\documentclass[journal,A4]{IEEEtran}

\usepackage{algorithm}
\usepackage[noend]{algorithmic}
\usepackage{epsfig}

\usepackage{subfigure}
\usepackage{psfrag}
\usepackage{threeparttable}
\usepackage{amssymb}
\usepackage[cmex10]{amsmath}
\usepackage{cite}
\usepackage{bm}

\newtheorem{theorem}{Theorem}
\newtheorem{lemma}{Lemma}
\newtheorem{definition}{Definition}

\newcommand{\nobreakhyphen}[2]{#1\nobreakdash-#2}
\newcommand{\refeqn}[1]{(\ref{#1})}
\newcommand{\reftable}[1]{Table~\ref{#1}}
\newcommand{\reffig}[1]{\figurename~\ref{#1}}
\newcommand{\refalg}[1]{Algorithm~\ref{#1}}

\newcommand{\reflemma}[1]{Lemma~\ref{#1}}
\newcommand{\refthm}[1]{Theorem~\ref{#1}}

\newcommand{\refsection}[1]{Section~\ref{#1}}

\newcommand{\remove}[1]{}
\def\TRUE{{\bf true}}
\def\FALSE{{\bf false}}

\begin{document}

\title{Stochastic Approximation Algorithm for Optimal Throughput Performance of Wireless LANs}
\author{Sundaresan Krishnan, Prasanna Chaporkar}
\maketitle

\begin{abstract}
Throughput improvement of the Wireless LANs has been a constant area of research.
Most of the work in this area, focuses on designing throughput optimal schemes for fully connected networks (no hidden nodes).
But, we demonstrate that the proposed schemes, though perform optimally in fully connected network, achieve significantly lesser throughput even than that of standard IEEE 802.11 in a network with hidden nodes.
This motivates the need for designing schemes that provide near optimal performance even when hidden nodes are present.
The primary reason for the failure of existing protocols in the presence of hidden nodes is that these protocols are based on the model developed by Bianchi in \cite{B00}. However this model does not hold when hidden nodes exist. Moreover, analyzing networks with hidden nodes is still an open problem. Thus, designing throughput optimal schemes in networks with hidden nodes is particularly challenging.
The novelty of our approach is that it is not based on any underlying mathematical model, rather it directly tunes the control variables so as to maximize the throughput. We demonstrate that this model independent approach achieves maximum throughput in networks with hidden terminals as well. Apart from this major contribution, we present stochastic approximation based algorithms for achieving weighted fairness in a connected networks. We also present a throughput optimal exponential backoff based random access algorithm. We demonstrate that the exponential backoff based scheme may outperform an optimal $p$-persistent scheme in networks with hidden terminals. This demonstrates the merit of exponential backoff based random access schemes which was deemed unnecessary by results in \cite{B00}.
\end{abstract}

\section{Introduction}
\label{Section:Introduction}
IEEE 802.11 has emerged as one of the most popular access mechanisms
for wireless local area networks (WLANs).
Since its inception, significant effort has been made for improving the throughput of IEEE 802.11~\cite{BFO96,HRGD05,CCG00_1, BCG04, HXHPSC04}.
Improving the throughput in WLANs is paramount in order to satisfy the ever increasing demand.
\par
In this paper, {\it we demonstrate that most of the proposed protocols though perform optimally for connected network (no hidden terminals), their throughput is lesser even than that of standard IEEE 802.11 in presence of hidden terminals.}
This motivates a need for designing schemes that not only perform well in connected networks, but also achieve near optimal throughput when hidden terminals are present. 
Our aim is to develop schemes to this end.
\par
At the heart of the IEEE 802.11 lies the Distributed Co-ordination Function (DCF)
based on CSMA/CA (Carrier Sense Multiple Access/Collision Avoidance).
DCF mechanism defines the channel contention mechanism and collision resolution scheme. Currently, DCF employs exponential backoff in which nodes cooperatively reduce their access probability upon collision.
In~\cite{B00}, mathematical modeling of the CSMA/CA protocol is carried out to quantify the throughput performance of the protocol.
In~\cite{B00}, the author has shown that the throughput offered by the IEEE 802.11 protocol
with standard parameter values is much less than the best and degrades significantly with increasing number of nodes. 
Another key finding of~\cite{B00} is: To maximize the system throughput, it is sufficient to choose a single backoff window appropriately for each terminal, and
one need not consider the exponential backoff.
This implies that it suffices to consider \nobreakhyphen{$p$}{persistent} CSMA rather than classical exponential backoff.
Most of the proposals aimed at maximizing the throughput of IEEE 802.11 DCF propose algorithms to tune the access probability of \nobreakhyphen{$p$}{persistent} CSMA adaptively by statistical estimations of the parameters.
For example,~\cite{BFO96, CCG00_2} estimate the number of nodes based on the number of busy slots observed in the system, and then choose appropriate access probability based on the obtained estimate.
Key observation used in computing the optimal access probability based on the estimated number of nodes is that the product of the number of nodes and the optimal access probability is approximately constant. 
In the recently proposed IdleSense algorithm~\cite{HRGD05}, the authors show that the number of idle slots remain constant when the access probability is optimally chosen. 
Thus, here the optimal access probability is obtained based on the estimated number of idle slots by each node.
\par
\begin{figure}[t]
    \centering
    \includegraphics[width=3in]{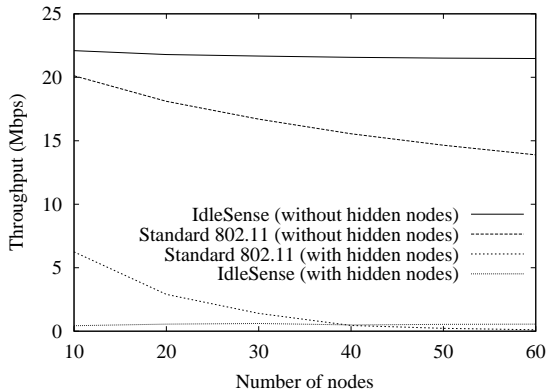}
    \caption{Comparison of throughput of standard 802.11 protocol and IdleSense protocol without and with hidden nodes. The simulation was performed in NS3 with nodes placed randomly and the results averaged over 20 iterations. The simulation parameters used are given in \reftable{Table:DCFParameters}. Hidden nodes are created by choosing the radius over which the nodes are distributed.}
    \label{Fig:IdleSense}
\end{figure}

The work described above uses analysis similar to that in~\cite{B00}, 
which holds only under fully connected network in which every node can perform carrier sensing on the
transmissions of every other node.
\emph{This model fails in the presence of hidden nodes.}
Node $i$ is hidden from node $j$ if $i$ is outside the sensing range of $j$, and as a result $i$ can not perform carrier sensing on $j$'s transmissions.
Hence, it is not clear how the proposed schemes that are optimal in fully connected network would perform in presence of hidden nodes.
To better understand the performance of the existing schemes in the presence of hidden nodes,
we perform the following simulation in {\it \nobreakhyphen{ns}{3}} network simulator.
We set parameters so that the transmission and sensing ranges are 16 and 24 units, respectively.
Hence, a node can potentially decode (sense, resp.) transmissions from the nodes at distance 
less than or equal to 16 (24, resp.) units from it.
We consider two types of network configurations,
first without hidden nodes and second with hidden nodes.
In the first case, nodes are placed uniformly on the edge of the disc with
radius 8 units centered at the access point.
In the second case, we place nodes uniformly at random in a disc of radius 
16 units centered at the access point.
 Note that the maximum distance between the
nodes is 32 units while the sensing radius is only 24 units, and 
hence there is a non-zero probability of having hidden node pairs.
In \reffig{Fig:IdleSense}, we compare the throughput of IdleSense protocol with that of standard 
IEEE 802.11 DCF with and without hidden nodes.
We choose IdleSense for comparison because it is the most recent protocol and has
been shown to outperform the other existing schemes.
From \reffig{Fig:IdleSense}, it can be seen that while IdleSense performs significantly 
better than the standard IEEE 802.11 DCF in the fully connected network,  but
the protocol performs much worse even when only a small number of hidden nodes exist.
The reason for the failure of the IdleSense like algorithms in the presence of hidden nodes can be understood as follows. 
These algorithms are based on the mathematical model in~\cite{B00} that govern the system. 
However when hidden nodes exist, the mathematical model does not hold.
For example the target number of idle slots, 
while known for a fully connected network, may be different 
for networks with different hidden nodes configurations. 
Hence it is not possible to design the system with 
any particular value of the target idle slots in presence of hidden nodes.
These observations motivate the need for schemes that not only perform
optimally in fully connected network but also provide near optimal throughput
in networks with hidden nodes.

Before proceeding further, the most important question that needs answering
is the following: {\it Do we really need to consider WLANs with hidden nodes?}
The question is important because mechanisms do exist to potentially eliminate hidden nodes.
For example, use of RTS/CTS exchange in IEEE 802.11 based MAC eliminates hidden nodes.
But this approach has its own limitations mainly because RTS and CTS messages are transmitted
at the lowest rate while data can be transmitted at a much higher rate.
For example, in IEEE 802.11a/g the maximum date rate is 54~Mbps while the RTS and CTS
messages are typically transmitted at 6 Mbps~\cite{802.11std}. Thus, overhead due to RTS/CTS exchange is
significant even though these control messages have much smaller length than that of
data packets. This affects throughput performance of WLAN adversely.
Hence, IEEE 802.11 standards recommend use of RTS/CTS exchange
only for large data packets. This is accomplished by choosing system parameter called
RTS threshold. If the length of data packet is above RTS threshold, only then RTS/CTS
exchange is used. By default, RTS threshold is set to the largest possible value, which is
2347 octets~\cite{802.11std}. Thus, the default setting ensures that the RTS/CTS exchange is not used.
Since, default settings are rarely changed during the installation in practice,
it is reasonable to consider WLAN operation in basic access mode, i.e.,
without RTS/CTS exchange.

Another approach to mitigate hidden nodes is by choosing sensing radius to be two times
transmission radius. Intuitively, since all the nodes have to be in the transmission range of access point,
with this setting, they have to be in the sensing range of each other.
Though this intuition may hold in open spaces, it may fail in spaces with obstacles, e.g. offices, universities.
This is because obstacles may cause strong shadowing between nodes.
Note that if the path between node $i$ and node $j$ is shadowed due to obstruction, then even though the receiver would be capable of decoding the data from both the nodes, the nodes will not be able to sense each other's transmissions.
Thus, this approach does not guarantee elimination of hidden nodes.
{\it This shows that hidden nodes can not be fully eliminated 
with current approaches. Hence, it is not sufficient
to only consider fully connected WLANs, rather we must consider
WLANs with hidden nodes.}

Now, lets understand challenges in designing throughput optimal schemes for WLANs with hidden nodes.
First note that
quantifying the throughput performance of IEEE 802.11 DCF in presence of hidden nodes has remained an open problem for
over a decade~\cite{TL08}.
To best of our knowledge, mathematical models for such networks are not available except for very 
restrictive special cases.
In absence of general mathematical models, obtaining provably throughput optimal schemes 
remained illusive. 
It is worth remembering that the existing model for the fully connected network
fail to provide any meaningful insight for the networks with hidden terminals.
In view of these difficulties, it is not clear how one should design schemes that
provide near optimal performance even in presence of hidden nodes.
In our approach, we do not base the design of optimal scheme on any
underlying model, rather we base it on the following simple observation:
If the channel access probability is ``small'', then the medium is underutilized 
resulting in low system throughput; on the other hand, if the channel access probability
is ``large'', then the system throughput is again low on account of excessive collisions.
This intuitively implies that the system throughput as a function of the access probability is bell-shaped
(technically quasi-concave).
Thus, we can employ standard gradient ascent techniques to obtain the optimal
access probability.
Now, we just need to verify that the system throughput is indeed a quasi-concave function
of the access probability. We prove the required for the fully connected networks,
which proves the optimality of our approach in this case.
However, in absence of a mathematical model for networks with hidden terminals, we
verify the quasi-concavity using extensive simulations in {\it \nobreakhyphen{ns}{3}}.
Of course, validation through simulations does not prove that our proposed scheme
is optimal when hidden terminals are present, but it at least states that in
the numerous random topologies that we investigated, our scheme is throughput optimal.

\remove{
\par
One work-around available in the IEEE 802.11 standards to eliminate hidden nodes is RTS/CTS mechanism that facilitates virtual carrier sensing.
However, it is required that the RTS and CTS packets be transmitted at the lowest rate which causes significant reduction in the system throughput because of the additional overhead of RTS/CTS exchange for each packet. 
Thus, eliminating hidden terminals using RTS/CTS mechanism may provide a lesser throughput than that in the system which does not use RTS/CTS~\cite{CBV04, SDB05}.
Hence, we can not fully eliminate hidden nodes without significant throughput penalty.}

Another interesting question that we address is: {\it How crucial exponential backoff
is in presence of hidden terminals?}
Note that in a fully connected network, the exponential backoff has shown to be unnecessary,
and optimizing attempt probability of a \nobreakhyphen{$p$}{persistent} scheme suffices in order to
maximize the system throughput~\cite{B00, BFO96}.
In this paper, we demonstrate that the exponential backoff based schemes can provide
a higher throughput than that under any \nobreakhyphen{$p$}{persistent} scheme.
Thus, \nobreakhyphen{$p$}{persistent} schemes may not be throughput optimal,
rather exponential backoff based schemes must be considered 
when hidden nodes exist. 

Our key contributions in detail are as follows:

\begin{itemize}
\item First, we consider a weighted fair throughput allocation while maximizing the system throughput. 
A weighted fair throughput allocation implies that the throughput allocated 
to a node is proportional to the weight assigned to the node.
Note that this is a generalized version of the system throughput maximization. 
When all weights are equal, weighted throughput maximization reduces
to the system throughput maximization.
We propose {\it Weighted Fair Throughput Optimal \nobreakhyphen{$p$}{Persistent} CSMA} ({\it \nobreakhyphen{$w$TOP}{CSMA}}) that is 
an on-line mechanism for tuning access probability of \nobreakhyphen{$p$}{persistent CSMA}. 
The tuning mechanism uses techniques from stochastic approximation theory~\cite{KY97}
that provides a weighted fair distribution of throughput in a connected network. 
In this algorithm every node achieves a throughput that is proportional to its weight. 
This algorithm does not require the nodes to know the weights used by the other nodes. 
Thus, every node could dynamically change their weights and the system would still adapt.

\item We also propose {\it Throughput Optimal \nobreakhyphen{RandomReset}{CSMA}} 
({\it \nobreakhyphen{TORA}{CSMA}}) that uses standard exponential backoff on transmission failure, 
but after successful transmission goes to a chosen backoff stage $j$ with probability (w.p.) $p_0$ 
and to any of the higher $m-j$ backoff stages w.p.~$(1-p_0)/(m-j)$. 
Again, \nobreakhyphen{TORA}{CSMA} uses stochastic approximation technique for on-line tuning of the parameter $p_0$ and $j$.

\item We show that both \nobreakhyphen{$w$TOP}{CSMA} and \nobreakhyphen{TORA}{CSMA} maximize system throughput while providing fairness, 
i.e. equal throughput to all nodes, in the connected network.

\item We also show that if the throughput is a quasi-concave function of the access probability 
in network with hidden nodes, then \nobreakhyphen{$w$TOP}{CSMA} maximizes the system 
throughput among all \nobreakhyphen{$p$}{persistent} CSMA schemes even when hidden nodes are present.

\item More interestingly, we show through simulations that \nobreakhyphen{TORA}{CSMA} may provide better throughput than that of 
\nobreakhyphen{$w$TOP}{CSMA} in presence of hidden terminals. 
This demonstrates the merit of exponential backoff when network is not fully connected. 
Thus, {\it \nobreakhyphen{$p$}{persistent} CSMA may not a good choice for networks with hidden terminals}.

\end{itemize}

\par
The rest of the paper is organized as follows.
\refsection{Section:SystemModel} explains the system model that is considered in this paper.
\refsection{Section:wTopCsma} provides the \nobreakhyphen{$w$TOP}{CSMA} policy and a proof of its optimality in the absence of hidden nodes.
\refsection{Section:ToraCsma} describes an exponential backoff based policy that guarantees optimal throughput allocation in a fully connected network when all nodes have equal weights.
\refsection{Section:OptimalPolicyWithHidden} discusses about the performance of the proposed policies in the presence of hidden nodes and conditions under which the policies are optimal.
\refsection{Section:Discussion} presents a discussion on the algorithms.
\refsection{Section:Simulation} presents the simulations results from {\it \nobreakhyphen{ns}{3}} simulations.
\refsection{Section:LittSurvey} lists the existing literature on the performance improvement of the 802.11 protocol and compares the results with our work.
\refsection{Section:Conclusion} provides conclusion.
\section{System Model}
\label{Section:SystemModel} 
We consider a system with $N$ nodes, and denote ${\cal N} = \{1,...,N\}$.
We consider the saturated case, i.e., all nodes always have a packet for transmission to a central Access Point (AP). 
Furthermore, when a node $t \in \mathcal N$ transmits, a subset ${\cal T}_t \subseteq \mathcal N$ of the nodes and AP receives the information.
We assume that $t \in {\cal T}_t$.
Note that $t$ is a hidden node for nodes in ${\cal N} \setminus {\cal T}_t$. We assume that all transmissions by AP
can be received by all nodes.
If ${\cal T}_t$ is same as the set ${\cal N}$ for every node $t$, then the network is called a {\it fully connected network}.
A node cannot receive and transmit simultaneously.
\par
For the channel access mechanism, we assume the following.
Any node $t \in {\cal N}$ waits for $CW$ slots  before attempting a transmission.
A slot can be either an {\it idle} slot of a pre-determined duration or a {\it busy} slot with some node transmitting. 
An idle slot for transmitter $t$ is a slot in which no transmission is received by $t$.
The duration of the idle slot is typically fixed by the standards, e.g., the IEEE 802.11 with the OFDM PHY on 20MHz channel has an idle slot duration of $9\mu s$.
A busy slot for $t$ is the duration for which it senses the channel to be busy followed by an idle duration of
DIFS (Distributed Inter-Frame Space).
Typically, DIFS is also defined by the standards, e.g., the IEEE 802.11 with the above PHY has DIFS equal to~$34\mu s$.
It is assumed that all nodes transmits the packet using the same fixed rate $R$.
\par
A transmission by $t$ to AP is successful if for the entire duration of transmission there is no other transmission by a node in ${\cal N}\setminus\{t\}$, i.e. the packet loss is only due to collisions \footnote{Though, we have not considered packet loss  due to channel errors, such can be incorporated in our framework in a straightforward fashion provided that the channel errors are independent and identically distributed over all transmissions.}. 
After a successful transmission, AP transmits an Acknowledgement (ACK) after a duration of SIFS (Short Inter-Frame Space).
According to IEEE 802.11 standard with the specified PHY SIFS is $16\mu s$.
If ACK is not received for DIFS duration after transmission (say by $t$), then $t$ concludes that the
transmission failed due to collision and performs a contention resolution algorithm.
The contention resolution algorithm specifies how $CW$ should be chosen for the next transmission.
Here, we consider three classes of contention resolution algorithms, namely, (1)~standard exponential backoff, (2)~\nobreakhyphen{$p$}{persistent}, and (3)~RandomReset.
Next, we briefly describe each of these.
\par
Standard exponential backoff mechanism refers to that used in IEEE 802.11.
Here, after $i$ successive transmission failures, $CW$ is chosen uniformly at random from the interval $[0, CW_{i}-1]$ where $CW_i = \min\{2^i\times CW_{\min}, CW_{\max}\}$.
While, after a successful transmission, $CW$ is chosen uniformly at random from the interval $[0, CW_{\min}-1]$. Here, $CW_{\min}$ and $CW_{\max}$ are the pre-defined system parameters.
The states in which $CW$ is chosen from different intervals are referred to as backoff stages.
Thus, the scheme has $m+1$ (= $\log_2(CW_{\max}/CW_{\min})$) backoff stages.
Typically, backoff stages are enumerated from 0 to $m$ with stage $i$ choosing $CW$ uniformly at random from $[0, CW_{i}-1]$.
\par
In \nobreakhyphen{$p$}{persistent} schemes, $CW$ is chosen as a geometrically distributed random variable with mean $1/p$ independent of success or failure of the previous transmission.
Here, $p$ is the system parameter.
Typically, $p$ is called the attempt probability, and the access mechanism is called \nobreakhyphen{$p$}{persistent} CSMA.
\par
In this paper, we propose RandomReset scheme. Broadly this scheme performs exponential backoff on transmission failures, but unlike
standard exponential backoff scheme, on successful transmission, it does {\it not} return to backoff stage 0 with probability (w.p.)~1.
Instead it probabilistically chooses a backoff stage with a probability distribution $\bm{q}$.
\par
In this paper, we consider both \nobreakhyphen{$p$}{persistent} and RandomReset mechanisms and show that their parameters $p$ and $\bm{q}$, respectively, can be tuned for optimal performance.
The performance metric that is primarily considered in the paper is {\it system throughput},
which is defined as the effective number of bits transferred to AP per unit time.
But, it is also important to consider notions of fairness.
Hence, we consider throughput optimization subject to fairness as defined below.
\par
\begin{definition}
A channel access scheme is said to be \emph{throughput optimal} if it maximizes
the system throughput while ensuring that each node has the same attempt
probability.
\end{definition}
\par
In this paper we also consider weighted fair throughput optimal schemes in which each node is pre-assigned with a weight indicating the priority of the node. 
A weighted fair throughput optimal scheme is defined as follows.
\par
\begin{definition}
A channel access scheme is said to be \emph{weighted fair throughput optimal} if it maximizes
the system throughput while ensuring that the throughput obtained by each node is proportional to its weight.
\end{definition}

In the following sections, we describe the proposed \nobreakhyphen{$w$TOP}{CSMA} and \nobreakhyphen{TORA}{CSMA}, and prove their throughput optimality in a fully connected network.
\section{A Weighted Fair Throughput Optimal policy in a fully connected network}
\label{Section:wTopCsma}
Here, we describe the Weighted fair Throughput Optimal \nobreakhyphen{$p$}{Persistent} CSMA algorithm (\nobreakhyphen{$w$TOP}{CSMA}).
We show that a weighted fair throughput allocation can be achieved by tuning a single variable. We then present an on-line algorithm to tune the variable to obtain optimal performance.
\subsection{Problem Formulation}
Let each node $t \in {\cal N}$ be assigned a fixed weight $w_t$ and denote $\bm{W} = [w_1, ..., w_N]$.
The problem of obtaining a weighted fair throughput optimal policy can be formulated as an optimization problem.
Let each station use a \nobreakhyphen{$p$}{persistent} CSMA.
Let $\bm{p} = [p_1, ..., p_N]$ be the vector of attempt probabilities with $p_t$ being the attempt probability of station $t$.
Let $S_t(\bm{p})$ be the throughput obtained by station $t$ when the attempt probabilities are $\bm{p}$. Let $S(\bm{p})$ be the system throughput.
Then the weighted fairness problem can be formulated by \refeqn{Eqn:WeightedFairOptimization}.

\begin{eqnarray}
\max_{\bm{p} \in [0, 1]^N} & C & \label{Eqn:WeightedFairOptimization}\\
\text{subject to} & \frac{S_t(\bm{p})}{w_t} = C & 1 \leq t \leq N \nonumber
\end{eqnarray}
Since system throughput maximization means maximization of $S(\bm{p}) = \sum_{t=1}^{N}S_t(\bm{p})=\sum_{t=1}^{N}w_tC$ which is same as maximizing $C$, this problem can be seen to maximize the system throughput.
Here $C$ is a constant which can be termed as the normalized throughput given to every station.
\par
This problem is a constrained optimization problem over an \nobreakhyphen{$N$}{dimensional} space.
In an adaptive setting this would require tuning of $N$ variables.
We show in the subsequent theorem that this problem can be reduced to a one-dimensional problem with a single control variable.
\par
The idea of a weighted fair throughput allocation has been discussed in~\cite{QS02} and~\cite{LJ08}. These works, while providing a weighted throughput allocation fail to provide an optimal throughput. In our work we consider a system throughput maximization problem under the constraint of a weighted throughput allocation.
\par
From the system definition it is possible to see that the throughput obtained by station $t$ is given by
\begin{equation}
S_t(\bm{p})=\frac{p_t}{1-p_t}\frac{E_P P_{I}}{P_{I}\sigma + P_{T}P_{I}(T_s-T_c) + (1-P_{I})T_c}.
\label{Eqn:AssymetricThroughput}
\end{equation}

Here $E_P$ is the expected packet length, $P_{I}=\prod_{i=1}^N(1-p_i)$ is the probability of an idle slot, $P_{T}P_{I}$ (with $P_{T}=\sum_{i=1}^N\frac{p_i}{1-p_i}$) is the probability that in a slot exactly one node transmits and $1-P_{I}-P_{T}P_{I}$ is the probability that more than one node transmits in the slot. $\sigma$ is the duration of an idle slot, $T_s$ is the duration of a slot with a successful transmission and $T_c$ is the duration of the slot with a failed transmission. In our model $T_s=(L_H+E_P)/R + T_{SIFS} + L_{ACK}/R + T_{DIFS}$ and $T_c=(L_H+E_P)/R+T_{DIFS}$, where $L_H$ is the length of the header of the MAC packet and $L_{ACK}$ is the length of the ACK packet. $T_{SIFS}$, is termed the short inter-frame space, is the duration after which the receiver responds with an ACK packet. $T_{DIFS}$ is the distributed inter-frame space which is the duration for which a contending node must sense the channel to be idle to identify the end of a transmission.
\par

\begin{lemma} \label{Lemma:WeightedFairRatio}
In a fully connected network using the \nobreakhyphen{$p$}{persistent} channel access scheme, if the attempt probability of station $i$ is $p_i$ and that of station $j$ is $p_j=\frac{wp_i}{1+(w-1)p_i}$ then $j$'s throughput equals $w$ times $i$'s throughput.
\end{lemma}
\par
{\it Remark:} Note that the relation between the throughput of $i$ and $j$ is independent of the attempt probability of the other stations in the system.
\begin{IEEEproof}
In the throughput equation for station $t$ given by \refeqn{Eqn:AssymetricThroughput} it can be seen that the values of $P_{I}$ and $P_{T}$ is the same for all stations.
Therefore $S_j(\bm{p})/S_i(\bm{p})$ is given by $\frac{p_j}{1-p_j}\frac{1-p_i}{p_i}$.
Equating this term to $w$, the value of $p_j$ in terms of $p_i$ can be found which gives the required equation.
\end{IEEEproof}
\par
If every station uses $p_t=w_{t}p/\left[1+(w_t-1)p\right]$, then the system throughput is given by
\begin{equation}
S(p, \bm{W})=\frac{E_P P_{T}P_{I}}{P_{I}\sigma + P_{T}P_{I}(T_s-T_c) + (1-P_{I})T_c}, \label{Eqn:WeightedFairSysThroughput}\\
\end{equation}
\begin{equation*}
\mbox{where}~P_{I}=\prod_{i=1}^{N}\frac{1-p}{1+(w_i-1)p}, P_{T}=\sum_{i=1}^{N}\frac{w_ip}{1-p}
\end{equation*}
\par
In the next theorem we show that the optimization problem in \refeqn{Eqn:WeightedFairOptimization} can be reduced to
\begin{equation}
\max_{p \in [0, 1]} S(p, \bm{W}) \label{Eqn:ModifiedWeightedFairOptimization}
\end{equation}

\begin{theorem}
The optimization problem in \refeqn{Eqn:WeightedFairOptimization} can be reduced to an unconstrained maximization over a single variable give in \refeqn{Eqn:ModifiedWeightedFairOptimization}.
\end{theorem}

\begin{IEEEproof}
From \reflemma{Lemma:WeightedFairRatio} it is seen that, for some arbitrary $p$, if the attempt probability of station $t$ is $p_t=\frac{w_tp}{1+(w_t-1)p}$ $\forall~t \in {\cal N}$ then station $t$ obtains a throughput $\frac{w_t}{\sum_{i=1}^{N}w_i}S$, where $S$ is the system throughput. {\it The attempt probabilities $p_t$ must necessarily satisfy this equation for obtaining a weighted fairness.} Hence the optimization problem reduces to \refeqn{Eqn:ModifiedWeightedFairOptimization}.
Here the system throughput $S(p, \bm{W})$ is given by \refeqn{Eqn:WeightedFairSysThroughput}.
Thus the \nobreakhyphen{$N$}{dimensional} optimization problem is reduced to a single variable problem.
\end{IEEEproof}
\par
From the above theorem it is seen that to obtain the weighted fair throughput optimal policy it is sufficient to tune a single variable.
We shall use an adaptive algorithm to tune the variable.
Before presenting the algorithm, we explain the intuition that led to the design of the algorithm.
Consider a fully connected network with $N$ nodes.
For this discussion we will assume that ever node is assigned a weight of $1$.
Bianchi has shown that there exists a unique access probability (say $p^\star$) for which the throughput is maximized~\cite{B00}.
Moreover, intuitively, it is clear that the throughput as a function of the access probability, denoted by $S(p, \bm{W})$ is monotone increasing for $p< p^\star$ and monotone decreasing for $p>p^\star$.
This is because for small $p$ the medium is under-utilized, while for large $p$ excessive collisions occur.
Thus, it appears that the optimal access probability can be obtained using some kind of gradient algorithm, where gradient is $\partial S(p, \bm{W})/\partial p$.
But, to calculate the gradient, we need value of $S(p, \bm{W})$ for a given $p$.
The throughput, in principle, can be computed at AP, and then AP can communicate the next value of $p$, computed using gradient ascent, to the nodes in ACK messages.
A key challenge in the aforementioned policy is that the exact value of the throughput can be obtained only through measurements for a long time as the underlying system is probabilistic.
Throughput measurement for a small duration can only yield noisy estimates.
Waiting for a long duration affects the convergence time of the algorithm.
Hence, we need techniques from stochastic approximation theory to circumvent the above problem.
Specifically, we use Kiefer-Wolfowitz algorithm that we succinctly describe next.

\subsection{Kiefer-Wolfowitz Algorithm}
In~\cite{KW52}, Kiefer and Wolfowitz have given an algorithm to obtain the maximum of a regression function using its stochastic estimates. 
Consider a function $S(p)$, and let the objective be to find the value of $p$ that maximizes $S(\cdot)$. 
However, the value of $S(p)$ is not known, but only a noisy estimate of it, namely $y(p)$ such that $\mathbb{E}[y|p] = S(p)$, is known.
Here $\mathbb{E}[\cdot]$ is the expectation operator.
Consider two sequences \{$a_k, k \geq 1$\} and \{$b_k, k \geq 1$\} satisfying 
$b_k \rightarrow 0$, $\sum_k a_k = \infty$, $\sum_k a_kb_k < \infty$, 
$\sum_k (a_k/b_k)^2 < \infty$, and the following recursive updates for $p$: 
\begin{equation}
    \label{Eqn:KW_Alg}
    p^{(k+1)} = p^{(k)} + a_k\frac{y_{2k} - y_{2k-1}}{b_k},
\end{equation}
where
$y_{2k}$ and $y_{2k-1}$ are noisy estimates of the function $S$ at $p^{(k)}+b_k$ and $p^{(k)}-b_k$, respectively.
Specifically $\mathbb{E}[y_{2k}] = S(p^{(k)}+b_k)$ and $\mathbb{E}[y_{2k-1}] = S(p^{(k)}-b_k)$. 
Then, the recursive scheme of updating $p$ given by \refeqn{Eqn:KW_Alg} converges stochastically to the optimal value $p^\star$ at which $S$ is a maximum provided $S(p)$ satisfies the following regularity conditions:

\begin{enumerate}
\item
$S(p)$ is a strictly quasi-concave function, i.e., 
there exists a $p=p^\star$ such that $S(p)$ is strictly increasing 
for $p<p^\star$ and $S(p)$ is strictly decreasing for $p>p^\star$.
  \item There exists $\beta$ and $B$ such that $|p_1-p^\star| + |p_2-p^\star| < \beta$ implies $|S(p_1)-S(p_2)| < B|p_1-p_2|$.
  \item There exists $\rho$ and $R$ such that $|p_1-p_2| < \rho$ implies $|S(p_1)-S(p_2)| < R$.
  \item For every $\delta > 0$ there exists a positive $\pi(\delta)$ such that $|p-p^\star| > \delta$ implies 
\begin{equation}
\inf_{0<\epsilon<\frac{\delta}{2}}\frac{|S(p+\epsilon) - S(p-\epsilon)|}{\epsilon} > \pi(\delta)\nonumber
\end{equation}
\end{enumerate}

The working of the algorithm can be understood by comparing it with the gradient ascent algorithm given as follows
\begin{equation*}
p^{(k+1)} = p^{(k)} + \alpha_{k}\nabla S(p^{(k)}),
\end{equation*}
where $\alpha_k$ is the step size.
This algorithm cannot be directly applied in our case,
since we do not know $S(p^{(k)})$ and hence $\nabla S(p^{(k)})$.
Hence, we wish to estimate the gradient through measurements.
To obtain an estimate of the gradient at $p^{(k)}$, we estimate the slope of the line
passing through points $(p^{(k)} - b_k, S(p^{(k)}-b_k))$ and $(p^{(k)} + b_k,S(p^{(k)}+b_k) )$
by measuring $S(p^{(k)}-b_k)$ and $S(p^{(k)}+b_k)$ (first order approximation of the gradient).
But, because of the stochastic nature of the system,
$S(p^{(k)}-b_k)$ and $S(p^{(k)}+b_k)$ can not be accurately obtained, rather only their
noisy estimates $y_{2k-1}$ and $y_{2k}$, respectively, are obtained.
Hence, the term $(y_{2k}-y_{2k-1})/b_k$ gives only the stochastic gradient 
of the function $S(\cdot)$ around $p^{(k)}$.
Now, to get close approximation of the gradient, one may take many
measurements and use their average. This is achieved by choosing $a_k/b_k \to 0$ as
$k\to\infty$ (ensured by $\sum_k (a_k/b_k)^2 < \infty$), which ensures that
adjustments in $p^{(k)}$ are much smaller than the gradient range $b_k$
as long as regularity conditions~2, 3 and~4 are met. 
Now, by choosing $b_k \to 0$ as $k\to\infty$,
the algorithm ensures that the evaluated gradient approaches the actual gradient.
Thus, the algorithm converges to $p^\star$ without knowing the exact form of $S(\cdot)$
whenever noisy estimates of the function can be obtained.
Next, we present adaptation of the Kiefer-Wolfowitz algorithm to our case.

\subsection{\nobreakhyphen{$w$TOP}{CSMA} Algorithm}
\begin{algorithm}[t]
\caption{\nobreakhyphen{$w$TOP}{CSMA} Algorithm to track the optimal value of the control variable $p$}
Algorithm at Access Point
\label{Algorithm:KW_Alg_Top}
\begin{algorithmic}[1]
\STATE $k \leftarrow 2$, $a_k \leftarrow 1/k$, $b_k\leftarrow1/k^{1/3}$
\STATE $p_{val} \leftarrow 0.5$, $p \leftarrow p_{val} + b_{k}$, $PlusSign \leftarrow \TRUE$
\IF{Packet is received successfully}
    \STATE $bytes\_recd \leftarrow bytes\_recd + \textbf{Packet\_Length}$
    \IF{$\textbf{Current\_Time} \geq last\_time+\textbf{UPDATE\_PERIOD}$}
        \IF{$PlusSign$}
            \STATE $S_{plus} \leftarrow bytes\_recd/\textbf{UPDATE\_PERIOD}$\label{AlgStep:ThroughputPlus}
            \STATE $PlusSign \leftarrow \FALSE$,  $p \leftarrow \max(p_{val} - b_{k}, 0)$
        \ELSE
            \STATE $S_{minus} \leftarrow bytes\_recd/\textbf{UPDATE\_PERIOD}$ \label{AlgStep:ThroughputMinus}
            \STATE $p_{val} \leftarrow p_{val} + a_{k} \frac{S_{plus}-S_{minus}}{b_{k}}$ \label{AlgStep:qUpdate}
            \STATE $k \leftarrow k+1$, $a_k \leftarrow 1/k$, $b_k\leftarrow1/k^{1/3}$
            \STATE $PlusSign \leftarrow \TRUE$,  $p \leftarrow \min(p_{val} + b_{k}, 0.9)$
        \ENDIF
        \STATE $bytes\_recd \leftarrow 0$, $last\_time \leftarrow \textbf{Current\_Time}$
    \ENDIF
    \STATE Transmit $p$ in the ACK packet
\ENDIF
\end{algorithmic}
Algorithm at Node $t$
\begin{algorithmic}[1]
\STATE $p_t \leftarrow 0.1$
\WHILE{Node in Network}
\STATE Transmit packet in a slot with probability $p_t$
\IF{ACK received}
  \STATE Obtain $p$ from the packet
  \STATE Set $p_t \leftarrow w_tp/[1+(w_t-1)p]$
\ENDIF
\ENDWHILE
\end{algorithmic}
\end{algorithm}

To obtain a weighted fair optimal \nobreakhyphen{$p$}{persistent} scheme, we need to tune the control variable $p$.
Each node can then use this control variable and the node's weight to arrive at the attempt probability that the node should use.
Tuning $p$ can be done using the pseudo code presented in \refalg{Algorithm:KW_Alg_Top}.
We term this algorithm \nobreakhyphen{$w$TOP}{CSMA} since this algorithm achieves a weighted fair throughput optimal system using the \nobreakhyphen{$p$}{persistent} scheme.
Next, we explain \refalg{Algorithm:KW_Alg_Top} in detail.
\par
The algorithm has one parameter: {\bf UPDATE\_PERIOD} duration 
(denoted henceforth as $\Delta$ for brevity), and it starts by
initializing variables $k=2$, $p=p_{val}+b_k$ and $p_{val}=0.5$.
For $p$ and $p_{val}$ any initial values between $0$ and $1$ is possible.
The time is divided into frames of duration $2\Delta$, and
each frame is subdivided into segments of duration $\Delta$.
The variables $k$ and $p_{val}$ are updated once at the end of each frame,
while variable $p$ is updated  once at the end of each segment.
In the first segment $p=p_{val}+b_k$, while in the second segment $p=p_{val}-b_k$.
AP transmits the value of $p$ in ACK messages.
Since all the nodes can hear ACKs transmitted by AP, they can
update their access probability using the value provided by AP and using their weight.
Thus a node with weight $w_t$ on receiving the value of $p$ in the ACK message sets the attempt probability to $p_t = w_tp/[1+(w_t-1)p]$.
The node then uses this updated attempt probability ($p_t$) to access the channel.
AP estimates throughput in each of the segments separately, where the throughput in the first (second, resp.) is denoted by $S_{plus}$ ($S_{minus}$, resp.).
Here $S_{plus}$ and $S_{minus}$ indicate the terms $y_{2k}$ and $y_{2k-1}$ in \refeqn{Eqn:KW_Alg}.
$(S_{plus}-S_{minus})/b_{k}$ is the stochastic gradient of the throughput function around the point $p_{val}$.
Using the stochastic gradient of the throughput around $p_{val}$, $p_{val}$ is updated towards the direction of gradient. 
This update is the same as that in \refeqn{Eqn:KW_Alg}.
We use $a_k = 1/k$ and $b_k = 1/k^{1/3}$. 
It may be noted that a small value of {\bf UPDATE\_PERIOD} will cause the estimated throughput to have a large variance hence causing a long time for convergence. On the other hand a large value of {\bf UPDATE\_PERIOD} will result in convergence in lesser iterations but still the convergence time would be large since the iterations run at a much lesser frequency. Hence it is required to choose the {\bf UPDATE\_PERIOD} appropriately. We find that an {\bf UPDATE\_PERIOD} that encompasses around 500 successful transmissions yields good results.

\subsection{Performance Guarantees of \nobreakhyphen{$w$TOP}{CSMA}}
It can be seen that the \nobreakhyphen{$w$TOP}{CSMA} converges to the optimal transmission probability provided that the throughput function satisfies the regularity conditions.
In the following theorem, we show that the throughput function indeed satisfies the regularity conditions.
First, let us define $T_s^\star$ and $T_c^\star$ as the expected duration of a
successful transmission and the expected duration of a failed transmission,
respectively measured in units of slot durations (i.e $T_s^\star = T_s/\sigma$ and $T_c^\star = T_c/\sigma$).

\begin{theorem} \label{Theorem:WeightedQuasi-concavity}
In a fully connected network, the throughput function of \nobreakhyphen{$w$TOP}{CSMA} $S(p, \bm{W})$ given by \refeqn{Eqn:WeightedFairSysThroughput} is a quasi-concave function of $p$ and satisfies the regularity conditions required by the Kiefer Wolfowitz Algorithm.
\end{theorem}

\begin{IEEEproof}
To show that $S(p, \bm{W})$ is quasi-concave function of $p$ we will show that there exists a $p^\star$ such that $S(p, \bm{W})$ is increasing for $p<p^\star$ and decreasing for $p>p^\star$. $S(p, \bm{W})$ can be re-written as 
\begin{eqnarray}
S(p, \bm{W}) &=& \frac{E_P}{\sigma(T_s^\star-T_c^\star + C(p, \bm{W}))}, \mbox{\ where}  \label{Eqn:CSMAThroughput} \\
C(p, \bm{W}) &=& \frac{T_c^\star}{P_TP_I} - \frac{T_c^\star-1}{P_T}
\end{eqnarray}
where $P_I=\prod_{i\in\mathcal N}(1-p_i)$, $P_T=\frac{p}{1-p}\sum_{i \in \mathcal N}w_i$ and $p_i=w_ip/\left[1+(w_i-1)p\right]$, $\forall~i \in \mathcal N$.
\par

By differentiating \refeqn{Eqn:CSMAThroughput} with respect to $p$, we obtain

\begin{eqnarray*}
    \frac{\partial S(p, \bm{W})}{\partial p}&=&-\frac{\sigma}{E_P}S^2(p, \bm{W})\frac{\partial C(p, \bm{W})}{\partial p}, \mbox{\ where} \\
    \frac{\partial C(p, \bm{W})}{\partial p}&=&-\frac{T_c^\star}{P_T^2P_I^2}(P_T\frac{\partial P_I}{\partial p} + P_I\frac{\partial P_T}{\partial p}) + \frac{T_c^\star-1}{P_T^2}\frac{\partial P_t}{\partial p},\\
\end{eqnarray*}
\begin{eqnarray*}
    \frac{\partial P_I}{\partial p}=-P_I\sum_{i \in \mathcal N}\frac{1}{1-p_i}\frac{\partial p_i}{\partial p} &, &\frac{\partial P_T}{\partial p}=\frac{1}{(1-p)^2}\sum_{i \in \mathcal N}w_i
\end{eqnarray*}
Thus $\partial C(p, \bm{W})/\partial p$ can be written as 
\begin{eqnarray*}
    \frac{\partial C(p, \bm{W})}{\partial p}&=&-\frac{\sum_{i \in \mathcal N}w_i}{(1-p)^2P_IP_T^2} f(p, \bm{W}), \mbox{where}\\
    f(p, \bm{W})&=& T_c^\star\left(1-p\sum_{i \in \mathcal N}\frac{1-p}{1-p_i}\frac{dp_i}{dp}\right) - (T_c^\star-1)P_I
\end{eqnarray*}

It is clear that $\partial S(p, \bm{W})/\partial p$ and $f(p, \bm{W})$ have the same signs. By substituting for $1-p_i$ and $dp_i/dp$ it can be seen that
\begin{equation*}
    f(p, \bm{W})= T_c^\star\left(1-\sum_{i \in \mathcal N}p_i - P_I\right) + P_I
\end{equation*}
Also
\begin{equation*}
    \frac{\partial f(p, \bm{W})}{\partial p}= -T_c^\star\sum_{i \in \mathcal N}\frac{\partial p_i}{\partial p}\left(1-\frac{P_I}{1-p_i}\right) -\sum_{i \in \mathcal N}\frac{\partial p_i}{\partial p}\frac{P_I}{1-p_i}
\end{equation*}
Since $\partial p_i/\partial p > 0$ and $0 \leq P_I/(1-p_i) \leq 1$, $\partial f(p, \bm{W})/\partial p < 0$.
Thus $f(p, \bm{W})$ is a monotonically decreasing function of $p$ with $f(0, \bm{W})=1$ and $f(1, \bm{W})=-(N-1)T_c^\star$.
Therefore, there exists a unique $p^\star \in \{p: 0 < p < 1\}$ such that $f(p^\star, \bm{W}) = 0$. This is the maxima of the function $S(p, \bm{W})$. 
Furthermore, $f(p)>0$ for $p<p^\star$ and $f(p, \bm{W})>0$ for $p>p^\star$. 
Thus, $\partial S(p, \bm{W})/\partial p > 0$ for $p<p^\star$ implying that $S(p, \bm{W})$ increases monotonically for $p<p^\star$. 
Similarly, $\partial S(p, \bm{W})/\partial p < 0$ for $p>p^\star$ implying that $S(p, \bm{W})$  decreases monotonically for $p>p^\star$. 
Hence, $S(p, \bm{W})$ is a strictly quasi-concave function in the range $0 < p < 1$. 

Furthermore, since $S(p, \bm{W})$, $C(p, \bm{W})$ and $f(p, \bm{W})$ are differentiable and bounded functions of $p$, it follows that $\partial S(p, \bm{W})/\partial p$ is also a continuous and bounded function of $p$ for $0 < p < 1$.
Thus, the other regularity conditions are also satisfied.
\end{IEEEproof}

\begin{figure}[t]
    \centering
    \includegraphics[width=3in]{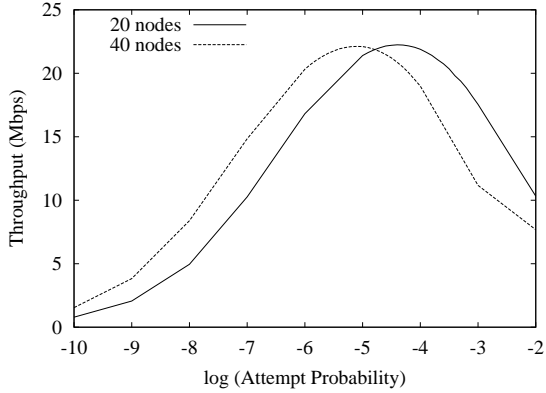}
    \caption{Throughput of the $p$-persistent CSMA versus the attempt probability}
    \label{Fig:pPersistentThrougputvsp0}
\end{figure}

\reffig{Fig:pPersistentThrougputvsp0} shows the bell shaped nature 
of the curve as proved in the above theorem. 
Note that
\begin{equation}
\label{Eqn:OptimalTau}
p^\star \approx \frac{1}{N\sqrt{T_c^\star/2}} 
\end{equation}
that is obtained in~\cite{B00} is an approximate solution to $\partial S(p, \bm{W})/\partial p=0$ when weights of all nodes are $1$.

\section{A Throughput Optimal exponential backoff policy in a fully connected network}
\label{Section:ToraCsma}
In a fully connected network, it has been shown that the exponential backoff based scheme
can be replaced with a \nobreakhyphen{$p$}{persistent} CSMA scheme without affecting the performance;
this is, however, not obvious in a network with hidden nodes. 
As seen from the simulations later, in a network with hidden nodes, it is possible that 
the optimal \nobreakhyphen{$p$}{persistent} scheme can perform worse even than the standard IEEE 802.11 protocol. 
This shows that an exponential backoff based scheme may have an advantage when hidden nodes exist. 
Hence, in this section we design an exponential backoff based scheme that can be tuned to achieve 
optimal performance in a fully connected network. 
We use this scheme instead of standard IEEE 802.11 DCF to better understand the merit of the
exponential backoff based schemes in presence of hidden terminals.
Since, our main motivation is only to assess the need of exponential backoff based schemes in networks with hidden terminals, for simplicity, we only consider
throughput optimal schemes, i.e. we assume that all nodes have the same weight.
\par
First, let us formalize our notion of the exponential backoff based schemes:
\begin{definition}
 An access mechanism is said to be exponential backoff based if 
(1)~the contention window size lies in $\{2^i CW_{\min}: i=0,\ldots,m\}$ for some
predefined parameters $CW_{\min}$ and $m$, and (2)~the contention window
size is doubled on a transmission failure except when the contention window size is at the
predefined maximum value, in which case it is kept the same.
\end{definition}
\par
Note that the exponential backoff based schemes are defined in terms of their
actions on transmission failures only, and their actions on transmission successes
are not specified. Thus, designing an optimal backoff based scheme amounts to
optimally choosing a backoff stage for a terminal when its transmission is
successful. We consider the following generic class of the exponential backoff
based schemes in which
each node uses a reset distribution denoted by $\bm{q}=[q_0, q_1, ..., q_m]$.
After a successful transmission, a node chooses contention window size $2^iCW_{\min}$ 
w.p.~$q_i$. 

{\it Remark:} Observe that the class we consider 
is {\it not} the class of all exponential backoff
based schemes.
One example of the exponential backoff based scheme that lies outside the
class that we consider is the following:
Each node maintains reset distribution $\bm{\tilde{q}_j}$ for $j=0,\ldots,m$, and
chooses backoff stage $i$ w.p.~$\tilde{q}_{ji}$ on transmission success given that
the node's current backoff stage is $i$. 
However, it suffices to consider this generic class as 
schemes in this class can achieve the same performance 
as any exponential backoff based scheme with the appropriate choice of $\bm{q}$.
This is because of the following observation:
In exponential backoff based schemes, the throughput of the system is 
determined by the node's attempt probability which is given by a 
fixed point solution~\cite{B00}.
Hence the performance of the system is defined by the attempt probability 
of the exponential backoff scheme for a given number of nodes.
It is intuitive that the smallest (largest, resp.) attempt probability is achieved 
if the node chooses stage $m$ ($0$, resp.) after every successful transmission.
Further, by using an appropriate probability distribution of choosing 
the backoff stage on a success any attempt probabilities in this range can be achieved.
Thus, the class of exponential backoff policies that we consider can achieve 
the same range of attempt probability as that by the entire class of the exponential backoff based policies,
 and hence the same throughput values.

Now, to obtain optimal exponential backoff scheme, we need to obtain appropriate
reset distribution $\bm{q}$. Thus, we need to tune $m+1$ parameters.
It is well known that convergence rate of on-line tuning schemes decreases rapidly as
the number of variables to be tuned increases. In our key contribution of the section,
we show that it suffices to tune only two parameters. To this end,
let us define RandomReset schemes.
\begin{definition}
A RandomReset($j$;$p_0$) is an exponential backoff policy with parameters
$j \in \{0,\ldots,m-1\}$ and $p_0 \in [0,1]$. 
The scheme chooses contention window $2^j\times CW_{\min}$
w.p.~$p_0$ and contention window $2^i\times CW_{\min}$ w.p.~$(1-p_0)/(m-j)$ 
for every $i\in\{j+1, \ldots, m\}$.
\end{definition}
\par
We term $p_0$ as the {\it reset probability}. We show that the optimal policy lies in the
domain of RandomReset schemes.
Designing optimal RandomReset policy requires tuning of two variables, 
namely, the backoff stage $j$ and the reset probability $p_0$. 
To design on-line algorithm for optimal tuning of RandomReset scheme, we use the following technique:
We fix $j$ and tune $p_0$ for RandomReset($j$;$p_0$) using Kiefer Wolfowitz algorithm.
This step is the same as that  in \nobreakhyphen{$w$TOP}{CSMA} (see \refalg{Algorithm:KW_Alg_Top}) except here
updates are applied to the reset probability $p_0$ instead of attempt probability $p$.
Now, if the value of $p_0$ drops below a fixed threshold $\delta_l (\approx 0)$, 
then we reset $j$ to $j+1$ and $p_0$ to $0.5$, except at boundary $j = m-1$. 
On the other hand, if $p_0$ rises above a fixed threshold $\delta_h (\approx 1)$, 
 then we reset $j$ to $j-1$ and $p_0$ to $0.5$, except at boundary $j=0$.
Even after reseting the parameters, we continue using Kiefer Wolfowitz algorithm
to tune $p_0$. Intuition behind the reset operation is as follows: Fix $j$.
If the maximum throughput is achieved in the range $0 < p_0 < 1$, 
then it implies that the throughput is optimal. 
However, if it is achieved at $p_0=0$ ($p_0=1$, resp.), then it implies that a better throughput 
can be achieved by increasing (decreasing, resp.) the backoff stage to decrease (increase, resp.) 
the attempt probability further. 
Hence by increasing or decreasing $j$ appropriately the system can be tuned 
until the optimal throughput is achieved. 
We term this algorithm the Throughput Optimal RandomReset CSMA algorithm (\nobreakhyphen{TORA}{CSMA} algorithm). 
\refalg{Algorithm:KW_Alg_Tora} gives the pseudo code for \nobreakhyphen{TORA}{CSMA}. Next, we establish optimality of \nobreakhyphen{TORA}{CSMA}.

\begin{theorem} \label{Theorem:OptimalRandomReset}
In a fully connected network, \nobreakhyphen{TORA}{CSMA} is optimal amongst the class of exponential backoff policies.
\end{theorem}

{\it Remark:} Though we claim the optimality of \nobreakhyphen{TORA}{CSMA} only in the class
of exponential backoff schemes, a more general result is true.
Specifically, it can be shown that there exist $N_l$ and $N_h$ such that
for every $N \in \{N_l,N_l+1,\ldots,N_h\}$ \nobreakhyphen{TORA}{CSMA} is throughput optimal
in the class of all policies. The bounds $N_l$ and $N_h$ on the number of nodes
in the system appears as the attempt probability under exponential backoff
policies is bounded below and above by $\tau_l > 0$ and $\tau_h < 1$, respectively.
This is because $CW_{min}$ and $m$ are fixed. Since for optimal throughput
the attempt probability should vary as $\Theta(1/N)$ with $N$ (see \refeqn{Eqn:OptimalTau}),
exponential backoff policies can not be optimal for every $N$.
Nonetheless, the range of $N$ for which \nobreakhyphen{TORA}{CSMA} is optimal among all policies is
significant. For example, for $CW_{\min}=8$ and $m=7$, $N_l=2$ and $N_h=140$, i.e \nobreakhyphen{TORA}{CSMA} is optimal for up to $140$ nodes.

\begin{algorithm}[t]
\caption{\nobreakhyphen{TORA}{CSMA} Algorithm to track the optimal value of $j$ and $p_0$}
Algorithm at Access Point
\label{Algorithm:KW_Alg_Tora}
\begin{algorithmic}[1]
\STATE $k \leftarrow 2$, $a_k \leftarrow 1/k$, $b_k\leftarrow1/k^{1/3}$, $j\leftarrow0$
\STATE $p_{val} \leftarrow 0.5$, $p_0 \leftarrow p_{val} + b_{k}$, $PlusSign \leftarrow \TRUE$
\IF{Packet is received successfully}
    \STATE $bytes\_recd \leftarrow bytes\_recd + \textbf{Packet\_Length}$
    \IF{$\textbf{Current\_Time} \geq last\_time+\textbf{UPDATE\_PERIOD}$}
        \IF{$PlusSign$}
            \STATE $S_{plus} \leftarrow bytes\_recd/\textbf{UPDATE\_PERIOD}$
            \STATE $PlusSign \leftarrow \FALSE$,  $p_0 \leftarrow \max(p_{val} - b_{k}, 0)$
        \ELSE
            \STATE $S_{minus} \leftarrow bytes\_recd/\textbf{UPDATE\_PERIOD}$
            \STATE $p_{val} \leftarrow p_{val} + a_{k} \frac{S_{plus}-S_{minus}}{b_{k}}$
            \IF{($p_{val} \leq \delta_l$) AND ($j \leq m-1$)}
              \STATE $p_{val} \leftarrow 0.5$, $j \leftarrow j+1$
            \ELSIF{($p_{val} \ge \delta_h$) AND ($j > 0$)}
              \STATE $p_{val} \leftarrow 0.5$, $j \leftarrow j-1$
            \ELSE
              \STATE $k \leftarrow k+1$
            \ENDIF
            \STATE $a_k \leftarrow 1/k$, $b_k \leftarrow 1/k^{1/3}$
            \STATE $PlusSign \leftarrow \TRUE$,  $p_0 \leftarrow \min(p_{val} + b_{k}, 1)$
        \ENDIF
        \STATE $bytes\_recd \leftarrow 0$, $last\_time \leftarrow \textbf{Current\_Time}$
    \ENDIF
    \STATE Transmit $p_0$, $2^j \times CW_{\min}$ in the ACK packet
\ENDIF
\end{algorithmic}
Algorithm at Node $t$
\begin{algorithmic}[1]
\STATE $p_0 \leftarrow 1$, $j \leftarrow 0$, $i\leftarrow 0$, $CW \leftarrow CW_{\min}$
\WHILE{Node in Network}
\STATE Transmit packet in a slot with probability $2/CW$
\IF{ACK received}
  \STATE Obtain $p_0$, $j$ from the packet
  \STATE Set $i \leftarrow j$ with probability $p_0$ and with probability $1-p_0$ choose $i$ uniformly between $j+1$ and $m$
\ELSE
  \STATE $i \leftarrow \min(i+1, m)$ 
\ENDIF
\STATE $CW \leftarrow 2^i \times CW_{\min}$
\ENDWHILE
\end{algorithmic}
\end{algorithm}

Two key steps in the proof of \refthm{Theorem:OptimalRandomReset} are:
(1)~We prove that the throughput function of the RandomReset($j$;$p_0$) policy 
is a quasi-concave function of $p_0$ for a fixed $j$.
(2)~We show that the RandomReset policy for $j$ varying from $0$ to $m-1$ 
can achieve the same attempt probability as that achieved by any general class of exponential based backoff schemes.
The detailed proof is presented in appendix.

\section{Performance of the Policies in the presence of hidden nodes}
\label{Section:OptimalPolicyWithHidden}

In a fully connected network a number of different algorithms have been proposed for throughput improvement. 
All these suggestions arrive at the required transmission probability that provides optimal throughput. 
If the number of stations in the system is known, then this transmission probability can be directly computed. 
However, this reasoning cannot be extended to the case of hidden nodes. 
This is because even if the number of stations are known beforehand, the number of possible configurations in which these stations could exist is large (exponential in $N$).
Here, configurations refer to various possibilities for ${\cal T}_t$ for each $t$.
It is required for the system to find out the exact configuration to perform optimization and arrive at a suitable transmission probability. 
This is clearly impractical, and hence the existing solutions cannot be used in such a scenario. 
For example, in the IdleSense protocol the target number of idle slots is a fixed constant.  
However, this target value changes in the presence of hidden nodes based on how many hidden node pairs exists. 
Hence, it is not possible to fix a target value of the idle slots a-priori.
\par
Analysis of hidden nodes has been an open problem for over a decade.
There has been different attempts at providing a mathematical model to analyze hidden nodes~\cite{TL08}.
However such attempts perform a simplification or consider restricted conditions.
Obtaining an analytical model for hidden nodes is a complicated task owing to various reasons.
One main hurdle is that in the presence of hidden nodes the system can no longer be modelled using an embedded Markov Chain.
The renewal equation used by Bianchi is no longer applicable.
Hence in~\cite{TL08} a fixed width slot model is studied for a system with two nodes hidden from each other.
In this work this model is shown for a system with two nodes.
However, this model becomes intractable even for small number of nodes (greater than 2).
\par
The lack of a mathematical model acts as an obstruction in the design of adaptive algorithms.
Algorithms that perform optimally in a fully connected network cannot be shown to perform optimally (or even better than standard 802.11) when hidden nodes are introduced.
In our work, while we do not eliminate this problem due to hidden nodes we show through extensive simulation that our algorithm can perform better than standard 802.11 even when hidden nodes exist.
The reason for this can be explained intuitively as follows.
The two optimal policies defined in \refsection{Section:wTopCsma} and \refsection{Section:ToraCsma}, namely, \nobreakhyphen{$w$TOP}{CSMA} and \nobreakhyphen{TORA}{CSMA}, are defined to track the actual throughput.
Specifically, these algorithms attempt to maximize the throughput directly using a stochastic approximation technique by taking a short term measurement of the actual throughput obtained from the system, and then dynamically tune the system parameters to increase the throughput.
The major advantage of these algorithms are that they have no binding on the mathematical model, but depend only on how the throughput varies as a function of the control variable.
As long as the throughput is a differentiable quasi-concave function of the control variable, the Kiefer Wolfowitz algorithm can obtain its optimal value. 
Unfortunately, it is not possible to prove analytically (even with a reasonable degree of approximation) that the function is differentiable and quasi-concave in presence of
hidden terminals.
Though a mathematical model does not exist when hidden nodes exist, it can be expected that the throughput is still a quasi-concave function of the attempt probability.
This is because for a small access probability, the throughput is small as the medium is under-utilized while for a large access probability, again the throughput is small 
because of excessive collisions. 
Thus, the throughput should increase until certain value of the access probability and decrease for the higher values, thereby yielding quasi-concavity.
Further in extreme situations where the throughput is not quasi-concave, the system would still tune the probability to one of the local maxima.
Hence it can be expected that the \nobreakhyphen{$w$TOP}{CSMA} and the \nobreakhyphen{TORA}{CSMA} algorithm perform better than standard 802.11.
We show this through simulations in {\it \nobreakhyphen{ns}{3}}. Also through simulation we obtain the plot of the throughput as a function of the attempt probability.
For the numerous random topologies that we simulated it was observed that the throughput is a quasi-concave function of the control variable.
Thus except for extreme cases, the proposed algorithms would perform reasonably well even when hidden nodes exists.

\label{Section:Discussion}
In this section, we discuss advantages and limitations of the proposed algorithms.
The \nobreakhyphen{$w$TOP}{CSMA} and \nobreakhyphen{TORA}{CSMA} algorithms are optimal in a fully connected network. 
The key feature of these algorithms that separates them from the existing proposals
is that these do not depend upon any underlying model, rather they act directly on the 
gradient of the throughput function. This feature makes them more robust towards scenarios in
which proposed mathematical models do not hold. 

Note that both the algorithms are centralized. 
While most of the existing algorithms are distributed algorithms, 
in our opinion, for infrastructure based WLANs implementation of the centralized algorithms would not pose many problems. 
Furthermore, having the intelligence in a centralized controller would facilitate 
easy modifications for performance improvements. 
Indeed, with the advent of IEEE~802.11e standard \cite{802.11estd}, the process of shifting responsibility of
smart, socially optimal decision making to the centralized controller has already begun.
Moreover, in a dynamic scenario in which nodes arrive and depart, the distributed algorithm may
not converge to the socially optimal point as the nodes already in the system will have to compete with
the new arrivals that tend to have very different initial values of control parameters. 

Now, our algorithms require that the parameters be passed in ACK packets. 
Though under \nobreakhyphen{TORA}{CSMA} it is sufficient for a node to process its own ACK,
under \nobreakhyphen{$w$TOP}{CSMA} a node has to process all the ACK packets which is not desirable. 
However, \nobreakhyphen{$w$TOP}{CSMA} can be modified to use beacon frames to send the parameters. 

Comparing the two algorithms, it can be seen that using a \nobreakhyphen{$p$}{persistent} CSMA provides the benefit of being able to achieve weighted fairness in a straightforward fashion. This is not easy in \nobreakhyphen{TORA}{CSMA}, since the throughput relation to the reset parameter $p_0$ is through a fixed-point solution and a closed form expression does not exist. In a fully connected network the \nobreakhyphen{$w$TOP}{CSMA} algorithm would suffice in providing an optimal throughput. However, it is not completely clear if this observation is true in a network with hidden nodes. As seen through simulations later, in a network with hidden nodes an exponential backoff scheme could provide better throughput. 
This shows that the exponential backoff schemes may not be eliminated all together.

\section{Simulation Results}
\label{Section:Simulation}
The IdleSense, \nobreakhyphen{$w$TOP}{CSMA} and \nobreakhyphen{TORA}{CSMA} algorithms are implemented in {\it \nobreakhyphen{ns}{3}} network simulator and their performance was compared without and with hidden nodes.
The parameters used in the simulations are given in \reftable{Table:DCFParameters}. The last two parameters, namely ns3::YansWifiPhy::EnergyDetectionThreshold and ns3::YansWifiPhy::CcaMode1Threshold ensure that nodes placed at a distance greater than 24~m do not sense the transmissions of each other. Such nodes behave as hidden nodes. 

\begin{table}[b]
\renewcommand{\arraystretch}{1.3}
\caption{Parameters used in the simulation (based on OFDM PHY with 20MHz channel
spacing~\cite{802.11std})} \label{Table:DCFParameters} \centering
\begin{tabular}{|l|l|}
\hline
Bit Rate & 54 Mbps\\
\hline
Packet Payload & 8000 bits\\
\hline
$CW_{min}$ & 8\\
$CW_{max}$ & 1024\\
\hline
ns3::YansWifiPhy::EnergyDetectionThreshold & -70 dBm\\
ns3::YansWifiPhy::CcaMode1Threshold & -70 dBm\\
\hline
\end{tabular}
\end{table}

In the simulation of \nobreakhyphen{$w$TOP}{CSMA} and \nobreakhyphen{TORA}{CSMA}, the {\textbf UPDATE\_PERIOD} was defined to be 250 ms. The IdleSense algorithm was implemented with a target average idle slots of 3.1 idle slots per transmission~\cite{HRGD05}.

\subsection{Performance of \nobreakhyphen{$w$TOP}{CSMA} in a fully connected network}
The simulation of \nobreakhyphen{$w$TOP}{CSMA} was carried out in a fully connected network by assigning weight to each node. The results of the simulation are shown for 10 nodes in \reftable{Table:wTopCsma}. As can be seen the normalized throughput obtained by every node is the same. Also a total throughput of 22 Mbps is obtained.

\begin{table}[b]
\renewcommand{\arraystretch}{1.3}
\caption{Simulation results of \nobreakhyphen{$w$TOP}{CSMA} algorithm}
\label{Table:wTopCsma} \centering
\begin{tabular}{|l|l|l|l|}
\hline
Node & Weight & Throughput & Normalized Throughput \\
 & & (Mbps) & (Throughput/Weight)\\
\hline
1  & 1 & 1.06624  & 1.06624 \\
2  & 1 & 1.06069  & 1.06069 \\
3  & 1 & 1.0603   & 1.0603  \\
4  & 2 & 2.17046  & 1.08523 \\
5  & 2 & 2.19471  & 1.09736 \\
6  & 2 & 2.11976  & 1.05988 \\
7  & 3 & 3.18249  & 1.06083 \\
8  & 3 & 3.18581  & 1.06194 \\
9  & 3 & 3.18676  & 1.06225 \\
10 & 3 &  3.19088 & 1.06363 \\
\hline
Total Throughput & \multicolumn{3}{c|}{22.4181 Mbps}\\
\hline
\end{tabular}
\end{table}

In the subsequent part, for comparison it is assumed that all nodes have equal weights.

\subsection{Comparison of performance of \nobreakhyphen{$w$TOP}{CSMA}, \nobreakhyphen{TORA}{CSMA} and IdleSense in a fully connected network}
\par
\begin{figure}[t]
    \centering
    \includegraphics[width=3in]{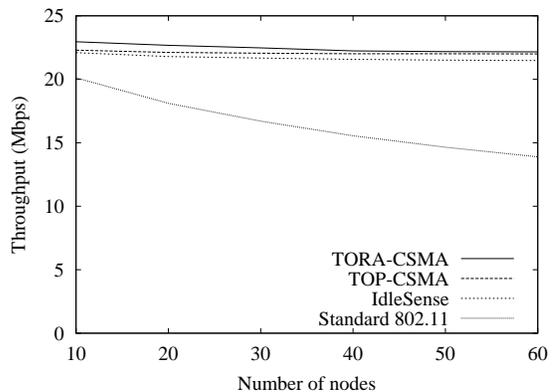}
    \caption{Comparison of throughput of standard 802.11 protocol, IdleSense and $w$TOP-CSMA and TORA-CSMA policies in fully connected network}
    \label{Fig:Performance_NoHidden}
\end{figure}

In \reffig{Fig:Performance_NoHidden} the throughput plots of the 
IEEE 802.11, IdleSense, \nobreakhyphen{$w$TOP}{CSMA} and \nobreakhyphen{TORA}{CSMA} as a function of the 
number of stations are shown. It is observed that \nobreakhyphen{$w$TOP}{CSMA} and \nobreakhyphen{TORA}{CSMA} perform much better than the Standard 802.11 protocol. Also the throughput obtained are as good as that obtained using IdleSense. However one advantage of the \nobreakhyphen{TORA}{CSMA} over the \nobreakhyphen{$w$TOP}{CSMA} can be seen by comparing \reffig{Fig:pPersistentThrougputvsp0} and \reffig{Fig:RandomResetThroughputvsp0}. The \nobreakhyphen{RandomReset}{CSMA} exhibits a more flat characteristics about the maxima while the \nobreakhyphen{$p$}{persistent} CSMA has a sharper fall from the maxima. This indicates that if the control variable oscillates around the optimal the throughput variations would be lesser for \nobreakhyphen{TORA}{CSMA} than that for \nobreakhyphen{$w$TOP}{CSMA}.

\subsection{Performance in the presence of hidden nodes}
We generate the networks with hidden nodes as follows: The nodes are created by randomly placing them in a disc of suitable radius with the Access Point at the center. The radius of the disc determines whether hidden nodes exist. Also, larger the radius more is the probability of hidden nodes. For the parameters that we configured in \nobreakhyphen{ns}{3} hidden nodes occur when the distance between the nodes is more than 24~m. Hence we consider the radius to be 16~m or 20~m for creating situations with hidden nodes scenarios.
\subsubsection{Quasi-Concavity of throughput}

\begin{figure}[b]
    \centering
    \includegraphics[width=3in]{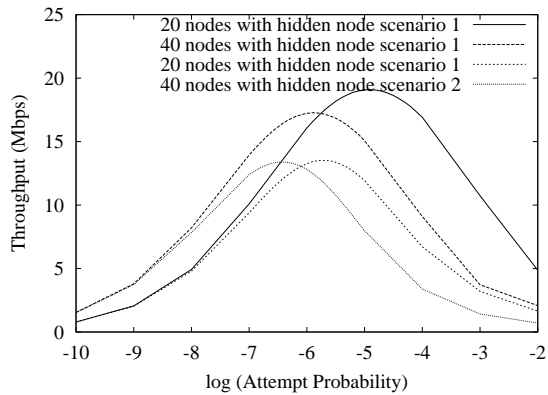}
    \caption{Throughput of the $p$-persistent CSMA versus the attempt probability in the presence of hidden nodes}
    \label{Fig:pPersistentThrougputvsp0_Hidden}
\end{figure}

\begin{figure}[t]
    \centering
    \psfrag{pzero}[][]{\Large{Probability of returning to Stage $0$ ($p_0$)}}
    \resizebox{3in}{!}{\includegraphics{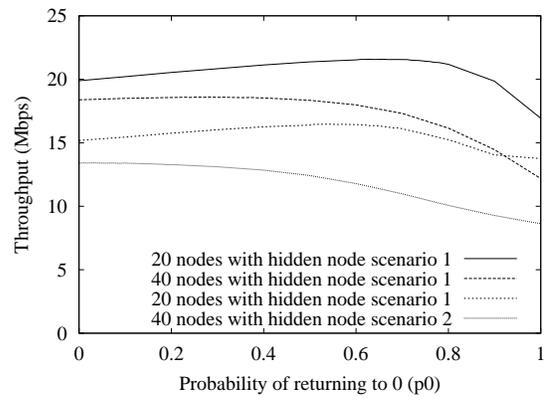}}
    \caption{Throughput of the RandomReset-CSMA policy versus $p_0$ in the presence of hidden nodes}
    \label{Fig:RandomResetThrougputvsp0_Hidden}
\end{figure}

For using the Kiefer Wolfowitz algorithm it was required that the throughput is a quasi-concave function of the control variable. For a network with hidden nodes, since this cannot be shown analytically, we present evidence of it through simulations. 
We verify the required by placing nodes randomly in a radius of 16~m or 20~m which creates a hidden node scenario. 
Subsequently, we compute the throughput for various values of the control variables in the generated network. 
In \reffig{Fig:pPersistentThrougputvsp0_Hidden}, the variation of the throughput of a \nobreakhyphen{$p$}{persistent} CSMA as a function of $p$ is shown, 
and in \reffig{Fig:RandomResetThrougputvsp0_Hidden}, the throughput of the RandomReset CSMA as a function of $p_0$ is shown. 
It can be seen that the throughput is indeed a quasi-concave function of the control variable. This characteristic has been noticed for a number of iterations of the simulation with different seeds for the random number generator.
This indicates that \refalg{Algorithm:KW_Alg_Top}, \refalg{Algorithm:KW_Alg_Tora} can be applied to obtain the optimal throughput. 

\subsubsection{Throughput Comparison}

\begin{figure}[b]
    \centering
    \includegraphics[width=3in]{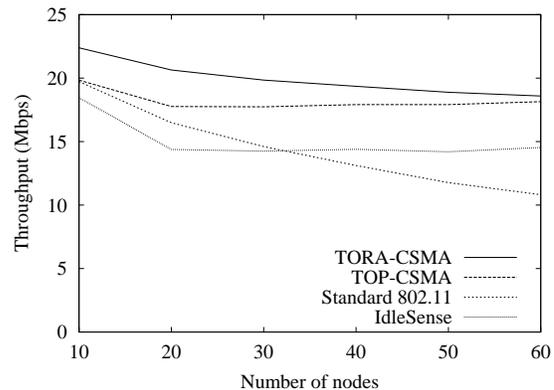}
    \caption{Comparison of throughput of standard 802.11 protocol, $w$TOP-CSMA and TORA-CSMA policies with nodes distributed in a disc of radius 16~m}
    \label{Fig:Performance_2_Hidden}
\end{figure}

\begin{figure}[b]
    \centering
    \includegraphics[width=3in]{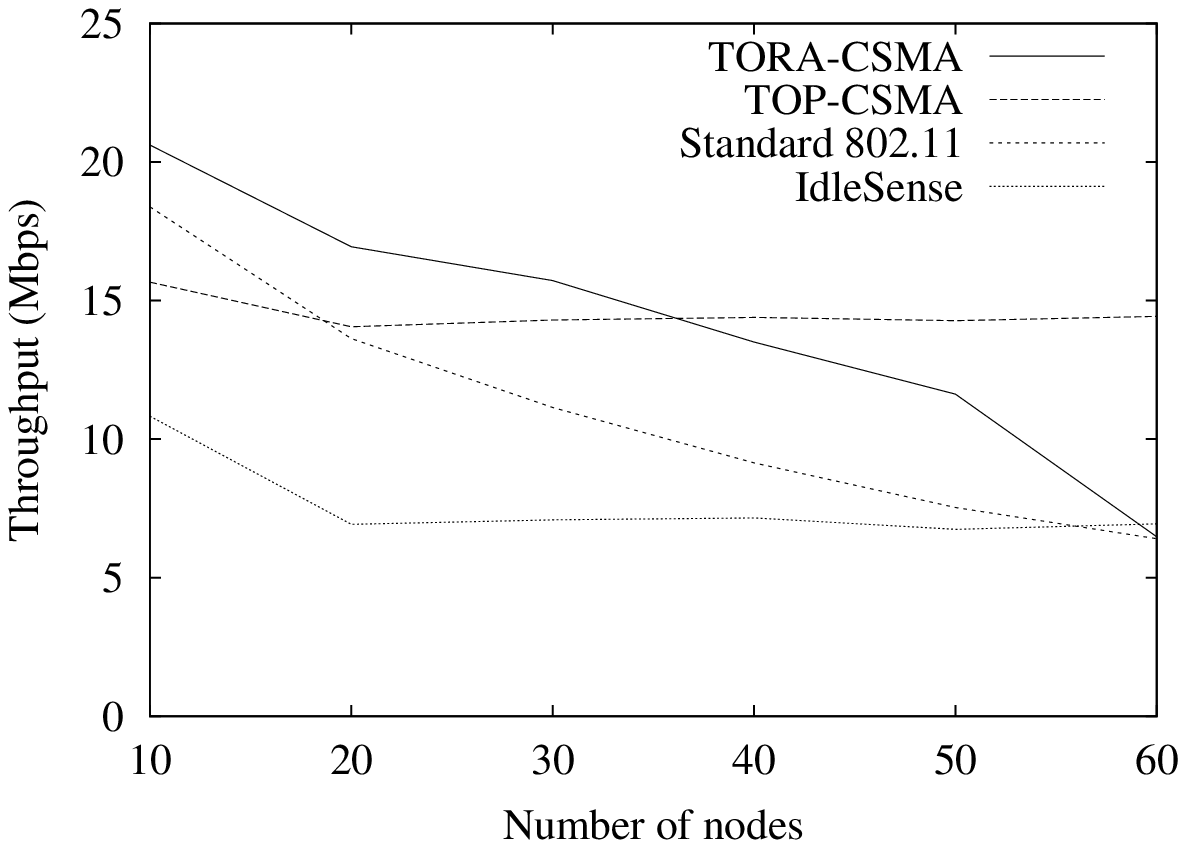}
    \caption{Comparison of throughput of standard 802.11 protocol, $w$TOP-CSMA and TORA-CSMA with nodes distributed in a disc of radius 20~m}
    \label{Fig:Performance_5_Hidden}
\end{figure}

The comparison of the performance of IEEE 802.11, \nobreakhyphen{$w$TOP}{CSMA} and \nobreakhyphen{TORA}{CSMA} 
is shown in \reffig{Fig:Performance_2_Hidden} and 
in \reffig{Fig:Performance_5_Hidden} for two scenarios. 
One key observation is that, unlike in the fully connected network, \nobreakhyphen{TORA}{CSMA} performs better than the \nobreakhyphen{$w$TOP}{CSMA} in the network  with hidden nodes. However as discussed in \refsection{Section:ToraCsma}, the \nobreakhyphen{TORA}{CSMA} algorithm may not be the optimal exponential backoff algorithm. The results show evidence that exponential backoff algorithm perform better than an optimal \nobreakhyphen{$p$}{persistent} channel access scheme when hidden nodes exist.
\par
From the simulation another inference that can be made is the reason for the failure of algorithms like IdleSense. In \reftable{Table:IdleSense} the idle slots obtained by the IdleSense algorithm and the \nobreakhyphen{$w$TOP}{CSMA} algorithm is shown. The first row shows the results when no hidden nodes are present while the second and third row show the results when hidden nodes are introduced. In each case IdleSense algorithm achieves a target average Idle slots per transmission value close to 3.1 as desired. However, from the results of \nobreakhyphen{$w$TOP}{CSMA} it can be seen that the target idle slots achieved is different for different scenarios. While it is close to 3 (4.8) when no hidden nodes exist, in the presence of hidden nodes it varies as 9 and 25. This shows that based on the configuration of the hidden nodes, the optimal idle slot per transmission value would vary. Hence it is not possible to choose a particular value when using the IdleSense algorithm. However, since our algorithm does not depend on system parameters, it still manages to track the optimal and achieve a relatively higher throughput. 

\begin{table}[!htb]
\caption{Comparison of average idle slots and throughput obtained for 40 nodes with and without hidden nodes by different schemes (Hidden nodes case 1 and case 2 refers to two runs of the simulation in {\it ns3})}

\label{Table:IdleSense}
\centering
\begin{tabular}{|c|c|c|}
\hline
    & \multicolumn{2}{c|}{IdleSense Algorithm}\\
\cline{2-3}
  & Average Idle Slots & Throughput (Mbps)\\
\hline
Without hidden nodes & 3.27849 & 21.5732 \\
With hidden nodes (case 1) & 3.29951 & 12.4798  \\
With hidden nodes (case 2) & 3.36866 & 0.497839 \\
\hline
\end{tabular}
\begin{tabular}{|c|c|c|}
\hline
    & \multicolumn{2}{c|}{\nobreakhyphen{$w$TOP}{CSMA} Algorithm}\\
\cline{2-3}
  & Average Idle Slots & Throughput (Mbps)\\
\hline
Without hidden nodes & 4.86102 & 22.0138 \\
With hidden nodes (case 1) & 9.97689 & 17.1832 \\
With hidden nodes (case 2) & 25.127 & 10.455 \\
\hline
\end{tabular}
\end{table}

\subsection{Dynamic Scenarios and Convergence Analysis}
\begin{figure}[t]
    \centering
    \includegraphics[width=3in]{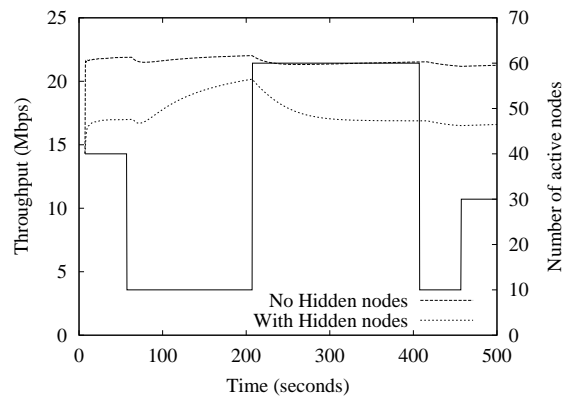}
    \caption{Variation of throughput of $w$TOP-CSMA as the number of stations vary}
    \label{Fig:Dynamism_Throughput_TOP}    
\end{figure}

\begin{figure}[t]
    \centering
    \includegraphics[width=3in]{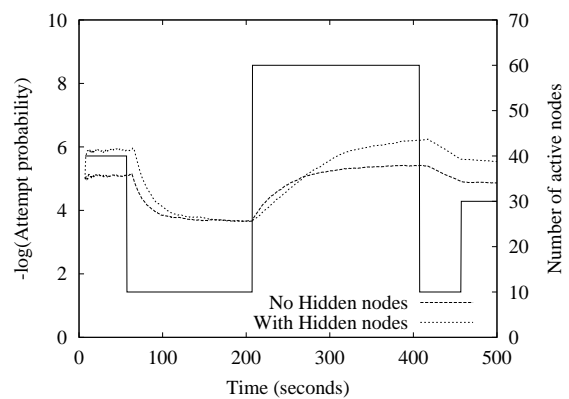}
    \caption{Variation of $p$ in $w$TOP-CSMA as the number of stations vary}
    \label{Fig:Dynamism_Probability_TOP}
\end{figure}

\begin{figure}[t]
    \centering
    \includegraphics[width=3in]{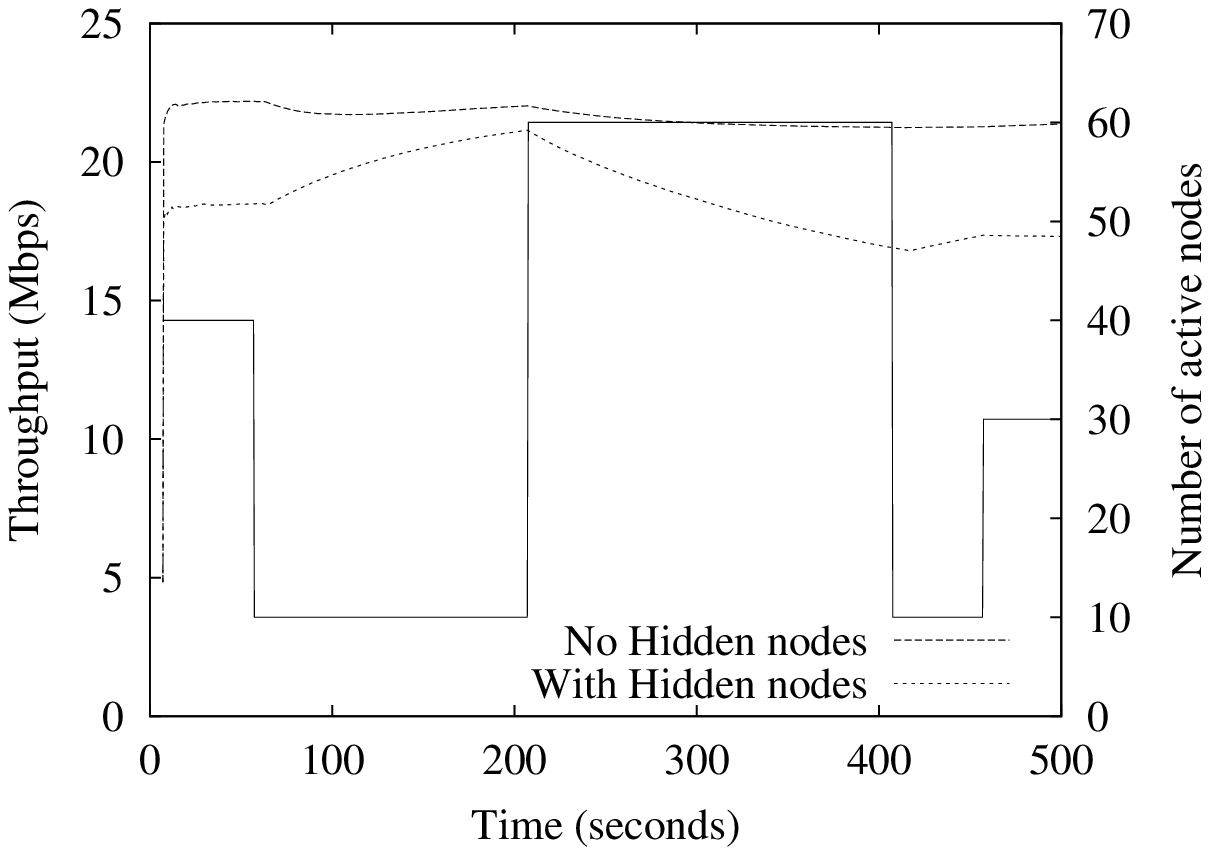}
    \caption{Variation of throughput of TORA-CSMA as the number of stations vary}
    \label{Fig:Dynamism_Throughput_Tora}
\end{figure}

\begin{figure}[t]
    \centering
    \psfrag{pzero}[][]{\Large{Probability of returning to Stage 0 ($p_0$)}}
    \resizebox{3in}{!}{\includegraphics{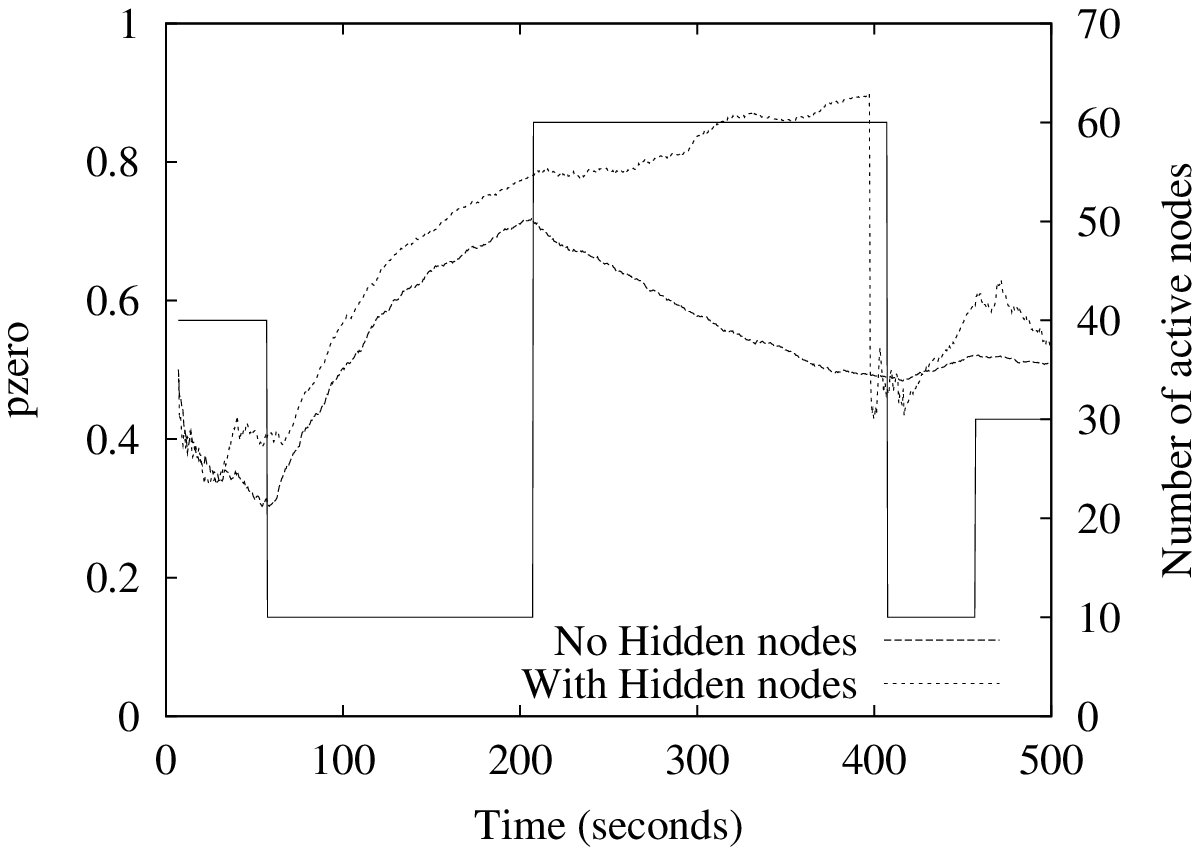}}
    \caption{Variation of $p_0$ of TORA-CSMA as the number of stations vary}
    \label{Fig:Dynamism_Probability_Tora}
\end{figure}

Another factor to be considered in the implementation of an adaptive protocol is the speed of convergence of the protocol, and the effectiveness of the protocol in changing conditions. 
In this case, the changes refer to the change in the number of active nodes $N$. 
We study the performance of \nobreakhyphen{$w$TOP}{CSMA} and \nobreakhyphen{TORA}{CSMA} algorithm when $N$ changes at predefined time instants. 
In \reffig{Fig:Dynamism_Throughput_TOP}, the variations in throughput as a function of time is shown (the number of nodes is indicated along the right y-axis) for \nobreakhyphen{$w$TOP}{CSMA} algorithm. 
Note that the throughput does not change very much even as $N$ changes when no hidden nodes exists. 
From \reffig{Fig:Dynamism_Probability_TOP}, it can also be seen that the system quickly adapts to changes and converges to the optimal probability value. 
The convergence for \nobreakhyphen{TORA}{CSMA} is shown in \reffig{Fig:Dynamism_Probability_Tora} and \reffig{Fig:Dynamism_Throughput_Tora}.
Thus, the algorithm suggested can handle dynamic conditions by modifying the control variable on-line as the number of nodes vary.

\section{Literature Survey}
\label{Section:LittSurvey}
There has been a number of attempts at designing algorithms that improve the throughput of IEEE 802.11. The key issue is the degradation of the throughput with increasing number of nodes. In this section we review the existing work on throughput improvement of IEEE 802.11.
\par
One of the first studies of the standard 802.11 was done in~\cite{B00} where the author provides a mathematical model to analyze the throughput of 802.11. The same author in~\cite{BFO96} also suggests an adaptive algorithm to improve the throughput of the protocol. The improvement is based on the idea that the optimal attempt probability is a function of number of nodes and hence by estimating the number of nodes the attempt probability can be chosen appropriately. However, the estimate of the number of nodes require the current value of the attempt probability and there is no way to ensure that every station uses the same attempt probability. For newly arriving stations the attempt probability being used is not known and hence it may cause unfairness to new nodes. Further, the optimal access probability depends on $T_c^\star$ which may vary based on the average packet length, the PHY layer used and the transmission rates.
\par
In~\cite{CCG00_1, CCG00_2} the authors again reduce the standard 802.11 to a \nobreakhyphen{$p$}{persistent} CSMA, and attempt to dynamically vary the attempt probability. The idea in this work is to keep the expected collision duration and the expected idle duration equal by tuning the attempt probability. To achieve this it is required to estimate the number of active stations which is done by measuring the average number of idle slots in a transmission time. However, this method does not work when hidden nodes are present as a proper estimate of the collision probability and average idle slots is no longer possible. The Asymptotic Optimal Backoff Mechanism defined in~\cite{BCG04} is an extension of the work in~\cite{CCG00_1, CCG00_2} where the optimal transmission probability is found by attempting to keep the slot utilization level at an optimum value called the Asymptotic Contention limit. This slot utilization level is achieved only asymptotically. Further this measure of the slot utilization cannot be done in the presence of hidden nodes.
\par
In~\cite{QABT03}, the authors suggest a slow Contention Window decrease rather than an immediate reset. This improves the throughput since the stations are less aggressive on success. However there is no guarantee of optimality of this protocol and as seen from the results presented in their work though there is an improvement over the standard 802.11 the throughput still degrades as a function of number of nodes.
\par
In~\cite{HXHPSC04}, a dynamic optimization is provided which estimates the approximate number of nodes and sets a contention window based on the range in which the number lies. This also uses a centralized AP based approach for optimization but requires an estimate of the number of nodes. The optimal contention window is transmitted in the beacon frame with which every station updates the $CW_{\min}$ value.
\par
Another improvement is the Idle Sense protocol given in~\cite{HRGD05}. IdleSense tunes the attempt probability so that the expected number of idle slots between transmissions remains a constant. Apart from throughput improvement the authors discuss about other factors like short term fairness and time fairness of the protocol. The short term fairness in the protocol is obtained as a result of reducing the binary backoff to a single \nobreakhyphen{$p$}{persistent} model. Further time fairness is an extension of the \nobreakhyphen{$p$}{persistent} CSMA model to provide a transmission probability that depends on the rate of transmission. Both these techniques can be extended to any \nobreakhyphen{$p$}{persistent} CSMA system including our system. However, as shown in our work the simple \nobreakhyphen{$p$}{persistent} CSMA cannot perform optimally in the presence of hidden nodes and hence a binary backoff is preferred. Furthermore in the Idle Sense protocol the optimal number of idle slots depends again on $T_c^*$ and varies based on the expected packet lengths and the PHY parameters. The optimal number of idle slots cannot be found out analytically in the presence of hidden nodes as there is no mathematical model defined. Further the optimal number of idle slots is expected to depend on the percentage of hidden nodes and cannot be dynamically found.
\par
In a recent work in~\cite{NA09} the authors suggest an improved backoff algorithm that looks at the history of channel utilization and uses a backoff that is non-binary. That is the algorithm attempts to increase the backoff window by $\alpha$ rather than a fixed value of 2. Like~\cite{QABT03}, though the protocol does perform better than IEEE 802.11, it does not perform optimally.
\par
The idea of a weighted fair throughput allocation, discussed in our work, has also been dealt with in other works. In~\cite{QS02} the authors arrive at the condition required for achieving weighted fair throughput allocation. Also the contention window to be used by each node for obtaining an optimum throughput is found. However the value of the optimal contention window depends on the number of nodes in each class. Thus every node in the network must be aware of the number of priority classes and number of nodes in each class. In~\cite{CCSC06} the authors consider only two classes of nodes. The contention window is arrived at numerically and hence requires an off-line calculation based on the number of nodes in each class. In~\cite{LJ08}, the authors suggest a modification to the algorithm to support weighted fairness. In this work while the throughput allocation is proportional to the weight of the node, it is not optimal. While it is shown that different attempt probabilities provide different throughput, no algorithm to achieve the optimal throughput is suggested. Unlike these works, our algorithm requires no knowledge of the number of nodes in different classes. The node is free to choose a weight independent of the other nodes and still obtain the optimal weighted fair allocation of the bandwidth.
\par
Recently in~\cite{RHCW09:1, RHCW09:2} the authors propose an algorithm to achieve \nobreakhyphen{$\alpha$}{fairness} for a general value of $\alpha$. Here, every node chooses an attempt probability that maximizes a certain utility function given the attempt probabilities of others. This process is done iteratively until the attempt probabilities converge. While the proposed algorithm provides good results, it also has some limitations. The algorithm requires that every node knows the attempt probability of every other node. This information can be obtained by message passing or through estimation. In a message passing technique, the number of messages passed for every iteration of the algorithm is of the order of the number of nodes in the system. Estimation, on the other hand, requires every node to process all packets in the network which is not desirable from a power consumption viewpoint. Also, the algorithm assumes that packet transmissions complete within a single slot. This assumption models a slotted Aloha kind of systems, and is far from an IEEE 802.11 based system in which packet transmission spans over multiple slots and CSMA is used.
\par
Another area of related study is the modeling of hidden nodes and the performance measurements in the presence of hidden nodes. Very few work exists that attempt to model the hidden nodes scenario and predict the system performance. One of the recent works in~\cite{TL08} show that existing modeling techniques would not work in the presence of hidden nodes and new approaches are required. The authors suggest a fixed slot technique and are able to model the system only when there are two nodes present. For a more general setup the system becomes too complicated to analyze mathematically. 
\section{Conclusion}
\label{Section:Conclusion}
In this work, we have suggested a modification to the CSMA/CA protocol of the IEEE 802.11 standard.
Unlike previous work, here the throughput is directly maximized by tuning a single control variable using stochastic maximization technique, specifically using the Kiefer Wolfowitz algorithm.
Two policies are suggested - one in which the control variable is the attempt probability of the p-Persistent CSMA and another in which the control variable is the probability of returning to a given backoff stage on success.
The modified algorithms $w$TOP-CSMA and TORA-CSMA are shown to perform optimally even in the presence of the hidden nodes.
It is seen that an exponential backoff based scheme may perform much better in the presence of hidden nodes as compared to the optimal $p$-Persistent CSMA.

\bibliographystyle{IEEEtran}
\bibliography{References,Fairness}

\begin{thebibliography}{10}
\providecommand{\url}[1]{#1}
\csname url@rmstyle\endcsname
\providecommand{\newblock}{\relax}
\providecommand{\bibinfo}[2]{#2}
\providecommand\BIBentrySTDinterwordspacing{\spaceskip=0pt\relax}
\providecommand\BIBentryALTinterwordstretchfactor{4}
\providecommand\BIBentryALTinterwordspacing{\spaceskip=\fontdimen2\font plus
\BIBentryALTinterwordstretchfactor\fontdimen3\font minus
  \fontdimen4\font\relax}
\providecommand\BIBforeignlanguage[2]{{%
\expandafter\ifx\csname l@#1\endcsname\relax
\typeout{** WARNING: IEEEtran.bst: No hyphenation pattern has been}%
\typeout{** loaded for the language `#1'. Using the pattern for}%
\typeout{** the default language instead.}%
\else
\language=\csname l@#1\endcsname
\fi
#2}}

\bibitem{B00}
G.~Bianchi, ``{Performance analysis of the IEEE 802.11 distributed coordination
  function},'' \emph{Selected Areas in Communications, IEEE Journal on},
  vol.~18, no.~3, pp. 535--547, Mar 2000.

\bibitem{BFO96}
G.~Bianchi, L.~Fratta, and M.~Oliveri, ``{Performance evaluation and
  enhancement of the CSMA/CA MAC protocol for 802.11 wireless LANs},'' in
  \emph{Personal, Indoor and Mobile Radio Communications, 1996. PIMRC'96.,
  Seventh IEEE International Symposium on}, vol.~2, Oct 1996, pp. 392--396
  vol.2.

\bibitem{HRGD05}
M.~Heusse, F.~Rousseau, R.~Guillier, and A.~Duda, ``{Idle sense: an optimal
  access method for high throughput and fairness in rate diverse wireless
  LANs},'' in \emph{SIGCOMM '05: Proceedings of the 2005 conference on
  Applications, technologies, architectures, and protocols for computer
  communications}.\hskip 1em plus 0.5em minus 0.4em\relax New York, NY, USA:
  ACM, 2005, pp. 121--132.

\bibitem{CCG00_1}
F.~Cal\`{\i}, M.~Conti, and E.~Gregori, ``{IEEE 802.11 protocol: design and
  performance evaluation of an adaptive backoff mechanism},'' \emph{Selected
  Areas in Communications, IEEE Journal on}, vol.~18, no.~9, pp. 1774--1786,
  Sep 2000.

\bibitem{BCG04}
L.~Bononi, M.~Conti, and E.~Gregori, ``{Runtime optimization of IEEE 802.11
  wireless LANs performance},'' \emph{Parallel and Distributed Systems, IEEE
  Transactions on}, vol.~15, no.~1, pp. 66--80, Jan. 2004.

\bibitem{HXHPSC04}
H.~Ma, X.~Li, H.~Li, P.~Zhang, S.~Luo, and C.~Yuan, ``{Dynamic optimization of
  IEEE 802.11 CSMA/CA based on the number of competing stations},'' in
  \emph{Communications, 2004 IEEE International Conference on}, vol.~1, June
  2004, pp. 191--195.

\bibitem{CCG00_2}
F.~Cal\`{\i}, M.~Conti, and E.~Gregori, ``{Dynamic tuning of the IEEE 802.11
  protocol to achieve a theoretical throughput limit},'' \emph{IEEE/ACM Trans.
  Netw.}, vol.~8, no.~6, pp. 785--799, 2000.

\bibitem{802.11std}
\emph{{IEEE 802.11 Part 11: Wireless LAN Medium Access Control (MAC) and
  Physical Layer (PHY) Specifications}}, IEEE, Mar. 2007.

\bibitem{TL08}
A.~Tsertou and D.~Laurenson, ``{Revisiting the Hidden Terminal Problem in a
  CSMA/CA Wireless Network},'' \emph{Mobile Computing, IEEE Transactions on},
  vol.~7, no.~7, pp. 817--831, July 2008.

\bibitem{KY97}
H.~J. Kushner and G.~Yin, \emph{{Stochastic Approximation Algorithms and
  Applications (Applications of Mathematics)}}.\hskip 1em plus 0.5em minus
  0.4em\relax Springer-Verlag Telos, January 1997.

\bibitem{QS02}
D.~Qiao and K.~Shin, ``Achieving efficient channel utilization and weighted
  fairness for data communications in ieee 802.11 wlan under the dcf,'' in
  \emph{Quality of Service, 2002. Tenth IEEE International Workshop on}, 2002,
  pp. 227--236.

\bibitem{LJ08}
J.-S. Liu, ``Achieving weighted fairness in ieee 802.11-based wlans: models and
  analysis,'' \emph{WTOC}, vol.~7, no.~6, pp. 605--615, 2008.

\bibitem{KW52}
\BIBentryALTinterwordspacing
J.~Kiefer and J.~Wolfowitz, ``{Stochastic Estimation of the Maximum of a
  Regression Function},'' \emph{The Annals of Mathematical Statistics},
  vol.~23, no.~3, pp. 462--466, 1952. [Online]. Available:
  \url{http://www.jstor.org/stable/2236690}
\BIBentrySTDinterwordspacing

\bibitem{802.11estd}
\emph{IEEE Standard for Information Technology - Telecommunications and
  Information Exchange Between Systems - Local and Metropolitan Area Networks -
  Specific Requirements Part 11: Wireless LAN Medium Access Control (MAC) and
  Physical Layer (PHY) Specifications Amendment 8: Medium Access Control (MAC)
  Quality of Service Enhancements}, IEEE, 2007.

\bibitem{QABT03}
Q.~Ni, I.~Aad, C.~Barakat, and T.~Turletti, ``{Modeling and analysis of slow CW
  decrease IEEE 802.11 WLAN},'' in \emph{Personal, Indoor and Mobile Radio
  Communications, 2003. PIMRC 2003. 14th IEEE Proceedings on}, vol.~2, Sept.
  2003, pp. 1717--1721 vol.2.

\bibitem{NA09}
Q.~Nasir and M.~Albalt, ``{Improved backoff algorithm for IEEE 802.11
  networks},'' in \emph{Networking, Sensing and Control, 2009. ICNSC '09.
  International Conference on}, March 2009, pp. 1--6.

\bibitem{CCSC06}
R.-G. Cheng, C.-J. Chang, C.-Y. Shih, and Y.-S. Chen, ``A new scheme to achieve
  weighted fairness for wlan supporting multimedia services,'' \emph{Wireless
  Communications, IEEE Transactions on}, vol.~5, no.~5, pp. 1095--1102, May
  2006.

\bibitem{RHCW09:1}
A.~Mohsenian-Rad, J.~Huang, M.~Chiang, and V.~Wong, ``Utility-optimal random
  access: Reduced complexity, fast convergence, and robust performance,''
  \emph{Wireless Communications, IEEE Transactions on}, vol.~8, no.~2, pp.
  898--911, Feb. 2009.

\bibitem{RHCW09:2}
A.~Rad, J.~Huang, M.~Chiang, and V.~Wong, ``Utility-optimal random access
  without message passing,'' \emph{Wireless Communications, IEEE Transactions
  on}, vol.~8, no.~3, pp. 1073--1079, March 2009.

\end{thebibliography}

\appendices

\section{Proof of \refthm{Theorem:OptimalRandomReset}}
First, we define some terms that are used in the proof. 
Let $\hat{\tau}(\bm{q})$ denote the attempt probability of nodes, which are using 
exponential back-off scheme with the reset distribution $\bm{q}$. 
For obtaining the attempt probability, we use the following assumption: 
the conditional probability of collision  is independent of the backoff stage of the nodes.
Note that the same assumption is used by Bianchi in his seminal paper~\cite{B00}.
Let $\hat{\tau}_{c}(\bm{q})$ denote the attempt probability 
for reset distribution $\bm{q}$ and the conditional collision probability $c$.
Similarly, let $\tau(j;p_0)$ and $\tau_c(j;p_0)$ denote the attempt probability
and the attempt probability given the conditional collision probability $c$, respectively,
under the RandomReset($j$;$p_0$) policy.
From the Markov Chain analysis of the system (similar to that in~\cite{B00}), 
the term $\hat{\tau}_{c}(\bm{q})$ is given by
\begin{eqnarray}
    \label{Eqn:GenericRandomResetTau}
    \hat{\tau}_c(\bm{q}) &=& \frac{\kappa_0}{\sum_{j=0}^{m}q_j\alpha_j(c)}, \mbox{\ where} \\
    \alpha_j(c)&=&(1-c)\sum_{i=j}^{m}2^ic^{i-j} + 2^m c^{m-j},~j=0, \ldots, m.\nonumber
\end{eqnarray}
Here $\kappa_j = 1/(2^{j-1}\times CW_{min})$ for $j=0,\ldots,m$. 
The conditional probability of collision is then given by
\begin{equation}
c = 1 - (1-\hat{\tau}_{c}(\bm{q}))^{N-1}.
\label{Eqn:CollisionProb}
\end{equation}
The term $\hat{\tau}(\bm{q})$ is then obtained as a fixed point solution of 
\refeqn{Eqn:GenericRandomResetTau} and \refeqn{Eqn:CollisionProb}.
The fixed point is unique \cite{B00}. Also, note from \refeqn{Eqn:GenericRandomResetTau},
the attempt probability given the conditional attempt probability $c$ for RandomReset$(j,p_0)$
is given as
\begin{eqnarray}
    \label{Eqn:RandomResetTau}
    \tau_c(j;p_0) &=& \frac{\kappa_0}{p_0\alpha_j(c) + \frac{1-p_0}{m-j}\sum_{i=j+1}^{m}\alpha_i(c)}.
\end{eqnarray}
\par
\begin{lemma}
\label{Lemma:Monotone}
Consider reset distributions $\bm{q}$ and $\bm{q'}$ such that
$\hat{\tau}_c(\bm{q}) < \hat{\tau}_c(\bm{q'})$ for every $c \in [0,1)$.
Then, $\hat{\tau}(\bm{q}) < \hat{\tau}(\bm{q'})$. 
\end{lemma}
\begin{IEEEproof}
 From \refeqn{Eqn:GenericRandomResetTau}, observe that $\hat{\tau}_{(\cdot)}(\bm{q})$
is a monotone decreasing and continuous function of $c \in [0,1]$ with range being a subset of $[0,1]$.
Thus, the graph corresponding to $\hat{\tau}_{(\cdot)}(\bm{q})$ lies below that corresponding to $\hat{\tau}_{(\cdot)}(\bm{q'})$.
Moreover, from \refeqn{Eqn:CollisionProb}, note that the conditional collision probability
$c$ as a function of attempt probability is a monotone increasing function and continuous with domain and range being $[0,1]$.
Thus, the unique point of intersection of graphs (fixed point of \refeqn{Eqn:GenericRandomResetTau} and \refeqn{Eqn:CollisionProb})
corresponding to $c$ and $\hat{\tau}_{(\cdot)}(\bm{q})$ lies to the left of that corresponding to $c$ and $\hat{\tau}_{(\cdot)}(\bm{q'})$.
Thus, the result follows.
\end{IEEEproof}
\begin{figure}
    \centering
    \psfragscanon
    \psfrag{p01}[Br][Br]{$p_0=0.0$}
    \psfrag{p02}[Br][Br]{$p_0=0.2$}
    \psfrag{p03}[Br][Br]{$p_0=0.4$}
    \psfrag{p04}[Br][Br]{$p_0=0.6$}
    \psfrag{p05}[Br][Br]{$p_0=0.8$}
    \psfrag{tauc1}[Br][Br]{$c=1-(1-\tau_c)^{N-1}$}
    \psfrag{xaxis}[][]{\Large{Conditional probability of collision ($c$)}}
    \psfrag{yaxis}[][]{\Large{Attempt probability $\tau_c(p_0;0)$}}
    \resizebox{3in}{!}{\includegraphics{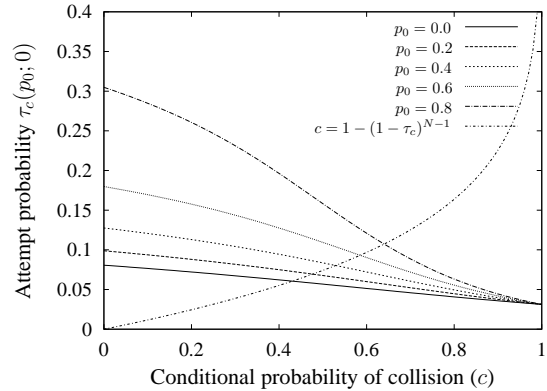}}
    \caption{Fixed point solution and monotonicity of attempt probability in TORA-CSMA for $N=10$, $m=5$, $CW_{min}=2$}
    \label{Fig:Monotonicp0}
\end{figure}

Following lemma, proves the uniform continuity of $\hat{\tau}(\cdot)$.
\begin{lemma} \label{Lemma:Continuity}
 The attempt probability $\hat{\tau}(\cdot)$ as a function of reset
distribution $\bm{q}$ is uniformly continuous, i.e.,
for every $\epsilon > 0$ there exists $\delta > 0$ such that 
\begin{equation*}
 \sup_{c\in [0,1]} |\hat{\tau}_c(\bm{q}) - \hat{\tau}_c(\bm{q'})| \le \delta \Rightarrow |\hat{\tau}(\bm{q}) - \hat{\tau}(\bm{q'})| \le \epsilon.
\end{equation*}

\end{lemma}
\begin{IEEEproof}
Let $c_{\bm{q}}$ be the conditional probability of collision when reset distribution is $\bm{q}$. That is $(c_{\bm{q}}, \hat{\tau}(\bm{q}))$ is the point of intersection of $\hat{\tau}_c(\bm{q})$ and \refeqn{Eqn:CollisionProb}.
Consider $\delta = \epsilon$. Let $\bm{q'}$ be such that
\begin{equation*}
\hat{\tau}_{c_{\bm{q}}}(\bm{q}) < \hat{\tau}_{c_{\bm{q}}}(\bm{q'}) \leq \hat{\tau}_{c_{\bm{q}}}(\bm{q}) + \epsilon.
\end{equation*}
Since \refeqn{Eqn:CollisionProb} is an monotonic increasing and continuous function of $c$ and $\hat{\tau}_{(\cdot)}(\bm{q'})$ is a monotonic decreasing and continuous function of $c$ the intersection of the two curves lies for $c > c_{\bm{q}}$ and for $\hat{\tau}$ in the range $(\hat{\tau}_{c_{\bm{q}}}(\bm{q}), \hat{\tau}_{c_{\bm{q}}}(\bm{q'}))$. Thus 
\begin{eqnarray*}
\hat{\tau}_{c_{\bm{q}}}(\bm{q}) & < \hat{\tau}(\bm{q'}) < & \hat{\tau}_{c_{\bm{q}}}(\bm{q'}) \leq \hat{\tau}_{c_{\bm{q}}}(\bm{q}) + \epsilon,\\
\implies \hat{\tau}(\bm{q}) & < \hat{\tau}(\bm{q'}) < & \hat{\tau}(\bm{q}) + \epsilon.
\end{eqnarray*}
For $\hat{\tau}_c(\bm{q'}) < \hat{\tau}_c(\bm{q})$ the above can be shown by replacing $\bm{q}$ and $\bm{q'}$. Combining both cases together we get the required result.
\end{IEEEproof}

The monotonicity property can be understood from \reffig{Fig:Monotonicp0}. From the plot it can be seen that since $\tau_c(p_0;0)$ is a monotonically increasing function of $p_0$ the curve for $\tau_c(p_0;0)$ for a higher value of $p_0$ lies above a curve corresponding to a lower value of $p_0$. Also the fixed point is the intersection of this curve with the curve for the plot of $c$ as a function of $\tau_c$. From the plot it can be seen that as $p_0$ increases intersection point shifts towards the top right indicating a higher attempt probability and a higher conditional collision probability.

In the following lemma, we establish some structural properties of the term in $\alpha_j(c)$ in \refeqn{Eqn:GenericRandomResetTau}.

\begin{lemma} \label{Lemma:TORABounds1}
For a given value of $c$ ($0 \leq c \leq 1$), $\alpha_1(c) \leq \ldots \leq \alpha_m(c)$
with equality only for $c=1$.
\end{lemma}
\begin{IEEEproof}
Note that $\alpha_j(c)$ can be written recursively as follows:
\begin{eqnarray*}
\alpha_m(c)&=&2^m,\\
\alpha_j(c)&=&(1-c)2^j + c\alpha_{j+1}(c), ~ 0 \leq j < m.
\end{eqnarray*}
First we show that $\alpha_{j}(c) \geq 2^j$ using backward induction. 
By definition this is true for $j=m$. 
Let this be true for $j=m,\ldots,k+1$. 
Now, we prove it for $j=k$.
\begin{eqnarray*}
\alpha_k(c)&=&(1-c)2^k + c\alpha_{k+1}(c)\\
           &\geq&(1-c)2^k + c2^{k+1} \ge 2^k,
\end{eqnarray*}
where the last step is because $c\geq0$.
Now, note that
\begin{eqnarray*}
\alpha_j(c)-\alpha_{j+1}(c)&=&(1-c)(2^j - \alpha_{j+1}(c)), ~ 0 \leq j < m\\
                           &\leq&(1-c)(2^j - 2^{j+1}) \le -2^j(1-c). 
\end{eqnarray*}
Thus, $\alpha_j(c)\leq\alpha_{j+1}(c)$ for  $0 \leq j < m$ with possible equality only for $c=1$.
\end{IEEEproof}
\par
Now, we establish monotonicity of the attempt probability under RandomReset policy.

\begin{lemma} \label{Lemma:MonotonicityTORA}
In a RandomReset($j$;$p_0$) policy the attempt probability of the nodes ($\tau(j;p_0)$) 
is a continuous and monotone increasing function of $p_0$ for a given $j$.
\end{lemma}

\begin{IEEEproof}
In view of \reflemma{Lemma:Monotone}, it suffices to show that
$\tau_c(j;p_0) < \tau_c(j;p_0^{\prime})$ whenever $p_0 < p_0^{\prime}$ for every $c \in [0,1)$.
We show this by showing that the partial derivative of $\tau_c(j;p_0)$ with respect to $p_0$
is positive for every $c \in [0,1)$.
Note that
\begin{eqnarray*}
\frac{\partial\tau_c(j;p_0)}{\partial p_0} &=& \frac{\tau^2_c(j;p_0)}{\kappa_0} \left\{\frac{1}{m-j}\sum_{i=j+1}^{m} \alpha_i(c) - \alpha_j(c)\right\}.
\end{eqnarray*}
Now, the monotonicity result follows by \reflemma{Lemma:TORABounds1} and continuity follows from ~\reflemma{Lemma:Continuity} since $\tau_c(j;p_0)$ is a continuous function of $p_0$.
\end{IEEEproof}
\par
Next, we show that the RandomReset($j$;$p_0$) achieves the entire range of attempt probabilities 
that can be achieved using the class of exponential back-off policies.

\begin{lemma}\label{Lemma:TORAtauBounds}
For any exponential back-off policy with reset distribution $\bm{q}$, $\hat{\tau}(\bm{q}) \in [\tau(m-1,0),\tau(0,1)]$.
\end{lemma}
\begin{IEEEproof} 
Note that for any $c \in [0,1)$
\begin{eqnarray*}
\hat{\tau}_c(\bm{q}) &=& \frac{\kappa_0}{\sum_{j=0}^{m}q_j\alpha_j(c)} \\
\Rightarrow \hat{\tau}_c(\bm{q}) &\in& \left[\frac{\kappa_0}{\alpha_m(c)},\frac{\kappa_0}{\alpha_0(c)}\right] \mbox{\ (from \reflemma{Lemma:TORABounds1})} \\
\Rightarrow \hat{\tau}_c(\bm{q}) &\in& [\tau_c(m-1;0),\tau_c(0,1)] \mbox{\ (from \refeqn{Eqn:RandomResetTau})}\\
\Rightarrow \ \hat{\tau}(\bm{q}) &\in& [\tau(m-1;0),\tau(0,1)] \mbox{\ (from \reflemma{Lemma:Monotone})}.
\end{eqnarray*}

This shows the required.
\end{IEEEproof}

\begin{lemma} \label{Lemma:RandomResetTauRange}
For every reset distribution $\bm{q}$, there exist $j$ and $p_0$ such that
$\hat{\tau}(\bm{q}) = \tau(j;p_0)$.
\end{lemma}
\begin{IEEEproof}
In view of \reflemma{Lemma:TORAtauBounds}, it suffices to show that
for any attempt probability in $[\tau(m-1;0),\tau(0,1)]$
can be achieved for some $j$ and $p_0$.

Note that for any $c$,
$\tau_c(j+1;1/(m-j)) = \tau_c(j;0)$. 
Thus, by \reflemma{Lemma:Monotone} and~\reflemma{Lemma:MonotonicityTORA},
\begin{equation} \label{Eq:Range1}
\tau(j+1;0) \leq \tau(j+1;1/(m-j))=\tau(j;0) \leq \tau(j+1;1).
\end{equation}
Further, since $\alpha_{j}(c) \leq \alpha_{j+1}(c)$ implies $\tau_c(j+1;1) \leq \tau_c(j;1)$,
we obtain from \reflemma{Lemma:Monotone} and~\refeqn{Eq:Range1}
\begin{equation*}
\tau(j+1;0) \leq \tau(j;0) \leq \tau(j+1;1) \leq \tau(j;1).
\end{equation*}

Now, the result follows from \reflemma{Lemma:MonotonicityTORA} and
\reflemma{Lemma:TORAtauBounds}.
\end{IEEEproof}

Now we prove that the throughput obtained by a RandomReset($j$;$p_0$) policy (denoted by $\tilde{S}(j,p_0)$)
is a quasi-concave function of the reset probability $p_0$ for a fixed $j$.
Note that $\tilde{S}(j,p_0) = S(\tau(j;p_0),\bm{1})$, where  $\bm{1}$ is a N-dimensional unit vector.

\begin{lemma} \label{Lemma:QuasiConcavityRandomReset}
The throughput function $\tilde{S}(j;p_0)$ (in the absence of hidden nodes) of a RandomReset($j$;$p_0$) policy 
is a strictly quasi-concave function of $p_0$ for a given $j$ and satisfies all the regularity conditions.
\end{lemma}
\begin{IEEEproof}
We know by \refthm{Theorem:WeightedQuasi-concavity} that $S(\tau(j;p_0),\bm{1})$ is a strict quasi-concave function 
of $\tau(j;p_0)$. Moreover, $\tau(j;p_0)$ is  
a monotone increasing function of $p_0$ (\reflemma{Lemma:MonotonicityTORA}).
Thus, $\tilde{S}(j,p_0)$ is also a quasi concave function of $p_0$. 
Also, $\tilde{S}(j,p_0) = S(\tau(j;p_0),\bm{1})$ satisfies the regularity conditions 
as shown in \refthm{Theorem:WeightedQuasi-concavity}.
\end{IEEEproof}

\begin{figure}[t]
    \centering
    \psfragscanon
    \psfrag{pzero}[][]{\Large{Reset probability ($p_0$)}}
    \resizebox{3in}{!}{\includegraphics{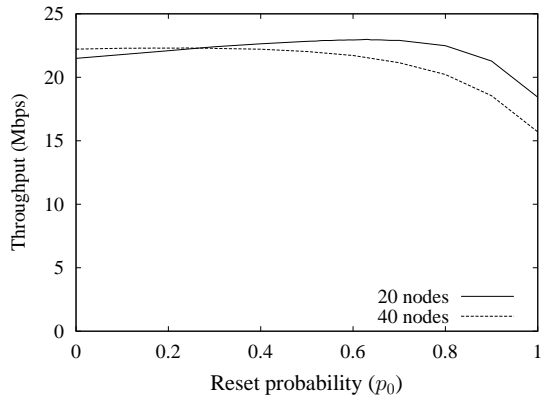}}
    \caption{Throughput of the RandomReset-CSMA policy versus $p_0$, for $j=0$}
    \label{Fig:RandomResetThroughputvsp0}
\end{figure}

The quasi-concavity can also be seen in \reffig{Fig:RandomResetThroughputvsp0}.

\begin{IEEEproof}[Proof of \refthm{Theorem:OptimalRandomReset}]
For a fixed $j$,
from \reflemma{Lemma:QuasiConcavityRandomReset}, we can conclude that the control variable ($p_0$) converges say to $p_0^{\star}$. 
There are two possibilities (1)~$p_0^{\star} \in (0,1)$, or (2)~$p_0^{\star} \in \{0,1\}$.
Let us consider case~(1). Since  $\tilde{S}(j,p_0)$ is a quasi-concave function of $p_0$,
$p_0^{\star}$ maximizes $\tilde{S}(j,p_0)$. Since $\tau(j;p_0)$ is a monotone increasing function of $p_0$ (\reflemma{Lemma:MonotonicityTORA}),
$\tau(j;p_0^\star)$ is at least a local maxima of $S(p,\bm{1})$.
But, since $S(p,\bm{1})$ is a quasi concave function of $p$, local maxima is also a global maxima.
Thus, RandomReset($j$;$p_0$) is throughput optimal policy.

Now, let us consider case~(2). $p_0^\star = 1$ (0, resp.) implies that 
 $\tilde{S}(j,p_0)$ is a monotone decreasing (increasing, resp.) function of
$p_0$. This in turn means that $S(p,\bm{1})$ is a monotone decreasing (increasing, resp.)
function for $p \in [\tau(j;0),\tau(j;1)]$.
Thus, optimal attempt probability is less than $\tau(j;0)$ (greater than $\tau(j;1)$, resp.).
To achieve this we need to increment (decrement, resp.) $j$ except when $j$ is at the boundary 
which is $j=m-1$ ($j=0$, resp.). If $j$ is at the boundary, we have optimal throughput in the class
of exponential back-off policies at $p_0^\star$. If $j$ is not on boundary with appropriate adjustment
of $j$ TORA-CSMA continues until it obtains $p_0^\star \in (0,1)$ for some $j$.
This proves the required.
\end{IEEEproof}

\end{document}